\documentclass[a4paper,aps,prb,twocolumn,floatfix]{revtex4}
\usepackage{graphicx} 
\usepackage{epsfig} 
\usepackage{amsmath}
\usepackage{amssymb} 
\newcommand{\be}{\begin{equation}}
\newcommand{\ee}{\end{equation}}
\newcommand{\sgn}[1]{\mathrm{sign}(#1)} 
\newcommand{\rr}{{\bf r}}
\newcommand{\qq}{{\bf q}} 
\newcommand{\kk}{{\bf k}}

\begin{document}
\title{Transport properties of a two-dimensional electron liquid at
high magnetic field} 
\author{Roberto D'Agosta}
\email{dagosta@fis.uniroma3.it} 
\affiliation{Dipartimento di Fisica,
Universit\`a di Roma ``Tre'', via della Vasca Navale, 84, 00146, Roma
Italy} 
\affiliation{Istituto Nazionale per la Fisica della Materia,
Unit\`a di Roma ``Tre''} 
\affiliation{NEST-INFM} 
\author{Roberto Raimondi}
\affiliation{Dipartimento di Fisica, Universit\`a di Roma ``Tre'', via
della Vasca Navale, 84, 00146, Roma Italy} 
\affiliation{Istituto
Nazionale per la Fisica della Materia, Unit\`a di Roma ``Tre''}
\affiliation{NEST-INFM} 
\author{Giovanni Vignale} 
\affiliation{Department
of Physics and Astronomy, University of Missouri Columbia, 65211,
Columbia, Missouri} 
\affiliation{NEST-INFM}

\begin{abstract}
The chiral Luttinger liquid model for the edge dynamics of a
two-dimensional electron gas in a strong magnetic field is derived
from coarse-graining and a lowest Landau level projection procedure at
arbitrary filling factors $\nu<1$ -- without reference to the quantum
Hall effect.  Based on this model, we develop a formalism to calculate
the Landauer-B\"uttiker conductances in generic experimental set-ups
including multiple leads and voltage probes.  In the absence of
tunneling between the edges the ``ideal" Hall conductances ($G_{ij}=
e^2 \nu/h$ if lead $j$ is immediately upstream of lead $i$,
and $G_{ij}=0$ otherwise) are recovered.  Tunneling of quasiparticles
of fractional charge $e^*$ between different edges is then included as
an additional term in the Hamiltonian.  In the limit of weak tunneling
we obtain explicit expressions for the corrections to the ideal
conductances.  As an illustration of the formalism we compute the
current- and temperature-dependent resistance $R_{xx}(I,T)$ of a
quantum point contact localized at the center of a gate-induced
constriction in a quantum Hall bar.  The exponent $\alpha$ in the
low-current relation $R_{xx}(I,0) \sim I^{\alpha -2}$ shows a
nontrivial dependence on the strength of the inter-edge interaction,
and its value changes as $e^*V_H$, where $V_H = h I/ \nu e^2$
is the Hall voltage, falls below a characteristic crossover energy
$\hbar c/d$, where $c$ is the edge wave velocity and $d$ is
the length of the constriction. The consequences of this crossover are
discussed vis-a-vis recent experiments in the weak tunneling regime.
\end{abstract}
\date{\today} 
\maketitle

\section{Introduction}
The Quantum Hall Effect (QHE) has been for the last twenty years an
amazingly rich source of experimental and theoretical results\cite{Klitzing1980, Tsui1982,Laughlin1983}
 (for a review see also Refs. \onlinecite{DasSarma1996, Chakraborty1988, Girvin1990}). 
Exotic concepts such as incompressible quantum Hall liquids,
fractionally charged quasiparticles\cite{Maasilta1997, Saminadayar1997,  de-Picciotto1997, Griffiths2000, Comforti2002} 
and Composite Fermions\cite{Jain1989, Heinonen1998} 
have become part of everyday's language of physics.  One of the
most interesting developments triggered by the QHE has been the
realization that the {\it edge} of a quantum Hall liquid\cite{Halperin1982}
provides a
clean realization of the chiral Luttinger Liquid
($\chi$LL)\cite{Wen1990,Wen1991a}.

As is well known, the Luttinger liquid (LL) concept -- first
introduced by Haldane\cite{Haldane1981}, building on an earlier exact solution
of the Luttinger liquid model\cite{Lieb1965, Luther1974} -- is the accepted paradigm for the
low-energy behavior of interacting Fermi liquids in one dimension.  In
the LL model two types of fermions -- right movers and left movers --
are coupled by an interaction of strength $g$.  Each type of fermion
by itself forms a chiral Fermi liquid, and its density fluctuation
$\delta \hat \rho_\alpha(x)$ ($\alpha$ = left (L) or right (R)) can be
expressed as the derivative of a bosonic displacement field $\hat
\phi_\alpha(x)$ that satisfies the commutation relations 
\be [\hat
\phi_\alpha(x),\hat \phi_\beta(x')] = i \pi s_\alpha \delta_{\alpha
\beta} \sgn{x-x'}~, 
\ee 
where $s_\alpha=1$ for $\alpha = L$ and
$s_\alpha=-1$ for $\alpha = R$.  The interaction between right and
left movers can be eliminated by a transformation (canonical up to a
scale factor) that preserves the relation between the net current and
the displacement fields and leads to two independent chiral fields
$\hat \phi'_\alpha$ which now satisfy the anomalous commutation
relation 
\be \label{anomalouscomm} 
[\hat \phi'_\alpha(x),\hat
\phi'_\beta(x')] = i \pi e^{-2 \theta} s_\alpha \delta_{\alpha \beta}
\sgn{x-x'}~, 
\ee 
where $\theta = \frac{1}{2} \tanh^{-1}\frac{g}{2}$ is
a measure of the strength of the original left-right coupling.  The
$\chi$LL model arises when one considers just {\it one} of these two
fields, with the commutator~(\ref{anomalouscomm}).

The anomalous commutator leads to a rich
phenomenology, including absence of the usual electron quasiparticles,
anomalously slow decay of correlation functions, nonlinear transport
properties etc...  Needless to say these effects are very difficult to
observe experimentally, due to the dramatic impact of even a modest
concentration of impurities on the properties of a one-dimensional
quantum system
\cite{Giamarchi1988,Kane1992,Kane1992b,Matveev1993,Yue1994, Furusaki1996,Lal2001,Kleinmann2002}.

It was therefore welcome news when, in a seminal 1990 paper,
Wen\cite{Wen1990,Wen1991a,Wen1995} showed that the density fluctuation
excitations at the edge of an incompressible quantum Hall liquid at
filling factor $\nu =1/q$ ($q$=odd integer) correspond to those of a
$\chi$LL with $e^{-2 \theta}=\nu$. Unlike one-dimensional metallic
systems, the edge of a quantum Hall liquid is
essentially unaffected by disorder, so the $\chi$LL ideas could
finally be put to an accurate experimental test\cite{Milliken1995, Chang1996}.
Following Wen's insight
the analysis was extended to more complex hierarchical QHE states,
where it turned out that one can have multiple branches of edge
excitations (i.e., multiple $\chi$LLs), some propagating in opposite
directions, and disorder plays a role in ensuring the correct
value of the quantized Hall conductance\cite{Kane1994,Kane1995}.

Understandably, these papers created a widespread belief that the
$\chi$LL behavior of the edge is inextricably tied to the QHE in the
bulk.  For one thing, the energy gap of the quantum Hall liquid state
was believed to be essential to ensure that the low energy excitations
are confined to the edges of the system.  It thus came as a big
surprise when Grayson et al. \cite{Grayson1998} reported that the
$\chi$LL could be observed in a whole range of filling factors
$\frac{1}{4}<\nu<1$ and was apparently unrelated to the quantization
of the Hall conductance.

In Section II of this paper we will argue that the validity of the
$\chi$LL model for the edge dynamics of a two-dimensional electron
liquid at high magnetic field follows from elementary semiclassical
considerations, which should be valid at any filling factor and have
nothing to do with the occurrence of the quantum Hall effect.  The
essential point is that the hydrodynamic modes, obtained by
``integrating out" fluctuations that are rapidly varying in time (on
the scale of the cyclotron frequency) and in space (on the scale of
the magnetic length) \cite{Aleiner1994}, are automatically bound to
the regions of space in which the gradient of the equilibrium density
differs from zero, i.e. to the edges of the system.  In addition, the
algebra of the edge density fluctuations (precisely defined in the
next section) follows from the algebra of the projected density
operators, when the latter is averaged on a length scale that is large
compared to the magnetic length.

In Sections III-IV we develop the formalism for the calculation of the
Landauer-B\"uttiker (LB)
conductances\cite{Datta1997} 
for generic experimental arrangements including
multiple terminals connected to the system by leads. Ordinarily, the
LB theory expresses the conductance $G_{ij}$ (which connects the
current in the $i$-th terminal to the voltage applied to the $j$-th
one) in terms of the transmission probability of an electron
quasiparticle from one terminal to the other.  But, in the present
case, there are no electron quasiparticles.  Instead, the voltage
applied to one terminal induces a train of collective waves which
propagate along the edges of the system and eventually feeds a current
into several different terminals.  It is not surprising therefore that
the conductance can be expressed solely in terms of the displacement
field propagators along the edges of the system.  We show that this
approach, in the absence of inter-edge coupling, yields the ideal Hall
conductances even in the presence of inhomogeneities that cause
partial reflection of edge waves.

Deviations from the ideal Hall effect can and do occur when the
possibility of inter-edge tunneling is taken into account.  This
subject is taken up in Sections V-VI.  Due to its quantum mechanical
origin tunneling is not included in the semiclassical hydrodynamic
description and must be introduced ``by hand".  We describe tunneling
in terms of two parameters, the tunneling amplitude $\Gamma$ and, most
importantly, the charge $e^*$ of the quasiparticles that are transferred
from one edge to the other.  Since a fundamental theory of $e^*$ at
general filling factors is not yet available one may choose to treat
$e^*$ as a phenomenological input parameter, whose value may be
determined from experiments.  Alternatively one can choose $e^* = \nu
e$ which is believed to be correct at $\nu = 1/q$, where $q$ is an odd
integer.  In terms of $\Gamma$ and $e^*$ we can finally calculate the
corrections to the ideal Hall conductances: the final expressions
involve the differential tunneling conductance, i.e. the derivative of
the tunneling current with respect to the potential difference between
the two edges.

 In Section VII we present a perturbative study of the nonlinear
 resistance $R_{xx}(I,T)$ of a quantum point contact situated within a
 constriction in a quantum Hall bar \cite{Roddaro2002}.  The
 perturbation theory is valid for $R_{xx} \ll \frac{h}{e^2}$.  This
 study generalizes Wen's original treatment of this phenomenon
 \cite{Wen1991b} and the later study by Moon and Girvin\cite{Moon1996} by
 including the effect of an inhomogeneous short-ranged inter-edge
 interaction, i.e., an interaction that is strong in the region of the
 quantum point contact, but becomes weak as one moves away from it.
 In Wen's paper a repulsive, but translationally invariant, inter-edge
 interaction leads to a {\it decrease} in the tunneling exponent
 $\alpha$ defined by $R_{xx}(I,0)\sim I^{\alpha -2}$ or
 $R_{xx}(0,T)\sim T^{\alpha -2}$.  This would make the behavior of the
 resistance even more singular than in the theory without inter-edge
 coupling at low temperature and bias voltage.  Our calculations
 indicate that the interplay of the inter-edge interaction with the
 broken translational invariance alters the relationship between the
 tunneling exponent and the strength of the interaction in the
 constriction region.  The new relationship is relevant when either
 $e^* V_H$, ($V_H = \frac{h I}{\nu e^2}$ being the Hall voltage) or
 $k_BT$ are above a geometric energy scale $\frac{\hbar c}{d}$, where
 $c$ is the edge wave velocity and $d$ is the length of the
 constriction.  For realistic values of the parameters this energy
 scale is in the range of $100$ mK.  Above this ``crossover" energy
 the tunneling exponent turns out to be larger than expected from the
 noninteracting theory and {\it a fortiori}, from Wen's interacting
 theory.

All these results suggest that a quantitative comparison between
theory and experiment cannot ignore the interactions between the edges
of the quantum Hall liquid in the region of the constriction.  In
particular the exponents of the current-voltage relationship may be
non-universal in the experimentally accessible range of temperatures,
reverting to universal values only at extremely low temperatures.
Evidence for nonuniversal behavior in the tunneling exponents has
recently surfaced from several different points of view
\cite{Shytov1998, Levitov2001, Mandal2002, Rosenow2002, Moore2002, Kane2003}.

\section{Derivation of the chiral Luttinger liquid model}
\label{general-problem}

Consider a two-dimensional electron liquid in a strong perpendicular
magnetic field ${\bf B} =- B {\bf \hat z}$ such that all the electrons
reside in the lowest Landau level (LLL).  The hamiltonian, projected
within the LLL, has the form
\begin{eqnarray}\label{hamiltonian}
\hat H &=&\frac12\int d\rr \int d\rr' \hat \rho(\rr) V(\rr-\rr') \hat
\rho(\rr') \nonumber \\ &&+~\int d\rr V_0(\rr) \hat \rho(\rr)~,
\end{eqnarray}
where $\hat \rho (\rr)$ is the number density operator {\it projected
in the LLL}, $V(\rr-\rr')$ is the electron-electron interaction
potential, and $V_0(\rr)$ is an external potential.  Both the kinetic
energy and a self-interaction-removing term are just constants, and
have therefore been dropped.

Next we write the density operator as the sum of the classical
equilibrium density $\rho_0(\rr)$ and a fluctuation $\delta \hat
\rho(\rr)$: 
\be 
\hat \rho(\rr) = \rho_0(\rr)+\delta \hat \rho(\rr)~,
\ee 
where $\rho_0(\rr)$ is determined by the equation 
\be \int d\rr'
V(\rr-\rr') \rho_0(\rr') + V_0(\rr) = \mu~, 
\ee 
and $\mu$ is a
constant fixing the total number of particles.  This gives (again, up
to a constant)
\begin{eqnarray}\label{hamiltonian2}
\hat H =\frac12 \int d\rr \int d\rr' \delta \hat \rho(\rr) V(\rr-\rr')
\delta \hat \rho(\rr')~.
\end{eqnarray}

The commutation relations between projected density operators at
different $\rr$ are easily deduced from the well known
result\cite{Girvin1986} 
\be 
\label{commutation1} [\hat \rho(\qq), \hat
\rho(\kk)] = \left (e^{k^*q \ell^2/2}- e^{-kq^*\ell^2/2} \right )\hat
\rho(\kk+\qq) , 
\ee 
where $k = k_x+ik_y$, $q=q_x+iq_y$, $\ell \equiv
\left (\frac{\hbar c}{eB} \right )^{1/2}$ is the magnetic length, and
$(x,y,z)$ form a right-handed coordinate frame.
 
Since we are interested in the dynamics of long wavelength density
fluctuations, we expand the right-hand side of ~(\ref{commutation1})
to leading order in $k \ell$, $q \ell$ and transform to real space.
This gives 
\be \label{commutation2} [\hat \rho(\rr), \hat \rho(\rr')]
\simeq i\ell^2 \epsilon_{ij}\partial_i \rho_0(\rr) \partial_j
\delta(\rr - \rr')~, \ee where $i, j$ denote cartesian components in
the $(x,y)$ plane, $\partial_i \equiv \frac{\partial}{\partial r_i}$,
$\epsilon_{ij}$ is the two-dimensional Levi-Civita tensor, and
repeated indices are summed over.  Notice that we have replaced the
density operator $\hat \rho(\rr)$ on the right hand side of
Eq.~(\ref{commutation2}) by its equilibrium expectation value
$\rho_0(\rr)$: this is legitimate as long as we are interested only in
the {\it linear dynamics} of small fluctuations about the equilibrium
state.

The remarkable feature of Eq.~(\ref{commutation2}) is that the
commutator is proportional to the derivative of the ground state
density.  This implies that hydrodynamic density fluctuations are
bound to regions where the equilibrium density varies, and are absent
from the regions of constant density.  This can be seen most clearly
by writing down the equation of motion for the density fluctuation,
which is easily seen to have the form \be \partial_t \delta\rho(\rr
,t) = \ell^2\left(\partial_{i}\rho_0(\rr)\right)
\varepsilon_{ij}\partial_{j}\int d\rr'V(\rr ,\rr') \delta\rho(\rr'
,t)~.  \ee Because this equation agrees with what one finds by taking
the large magnetic field limit of the hydrodynamic Euler
equations\cite{Aleiner1994} we will call our approach
``hydrodynamical".

\begin{figure}[ht]
\includegraphics[clip,width=7cm]{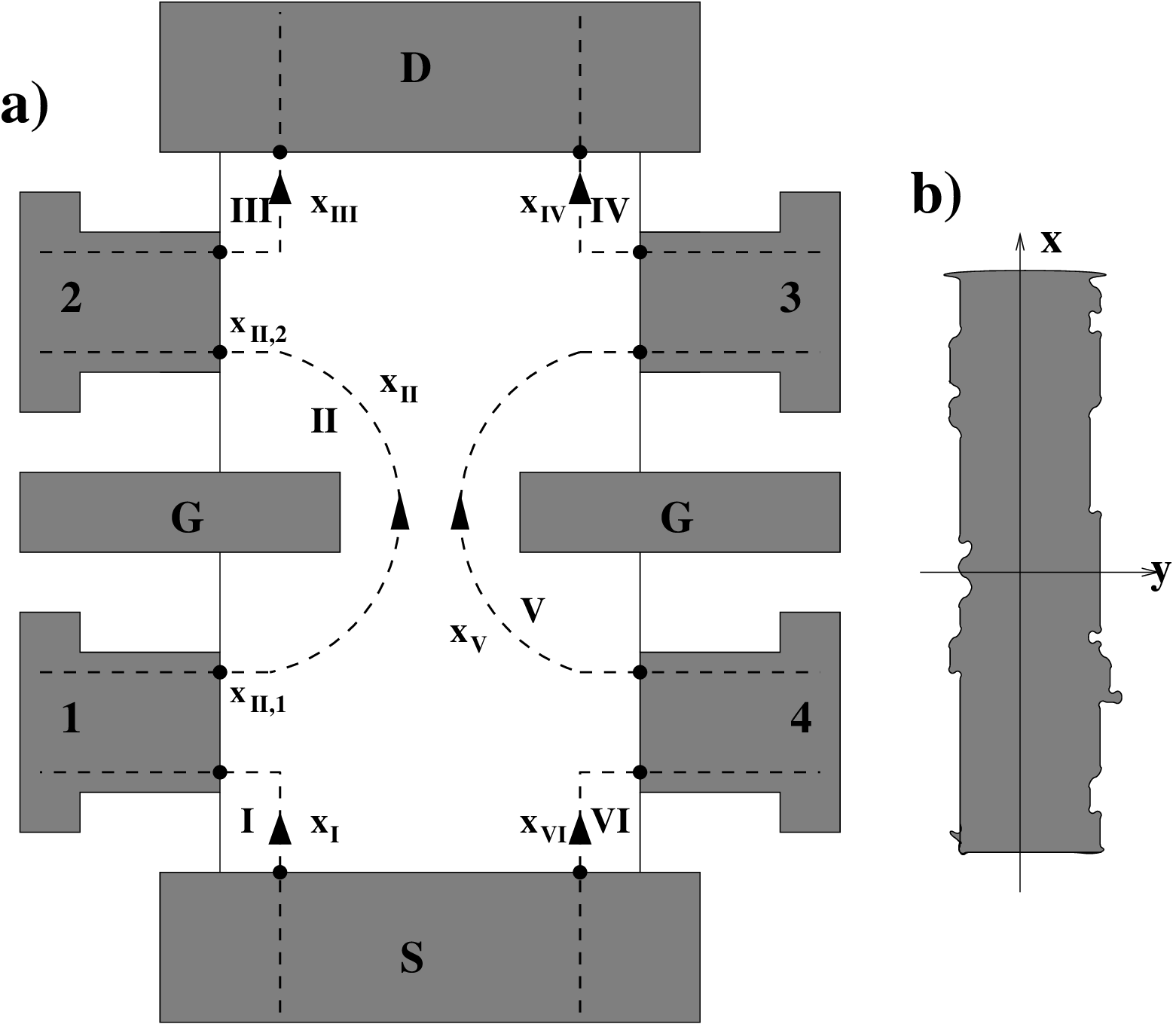}
\caption{a) Description of the device used in the experiments. A
two-dimensional electron gas is in contact with several
reservoirs. Two, the source (S) and the drain (D), are used to inject
and extract a current. The others (four in the figure) are used as
voltage probes.  Each reservoir contacts two edges (labelled by roman
numerals). Gate voltages (G) create a depletion zone and force the 
edges to stay close. The current flows are shown in the figure.  b) Expanded
view of the edge region.  Density fluctuations exist only in a limited
region around the $x$ axis (the shaded region in the figure).  We
integrate over the direction $y$ perpendicular to the edge to obtain
an ``edge density fluctuation" that depends only on the position along
the edge.}
\label{fig_probe.eps}
\end{figure}

The fact that the hydrodynamic density fluctuations are proportional
to the derivative of the equilibrium density implies that they are
concentrated near the edges of the system where the density profile
has a strong variation.  Let us consider, for definiteness, the model
depicted in Fig. \ref{fig_probe.eps}, where the boundary of the
electron liquid is divided by leads into different ``edges" labelled
by roman numerals.  The leads are connected to reservoirs (labelled
1--4 in Fig. \ref{fig_probe.eps}).  There are also two special
reservoirs, the ``source" (S) and the ``drain" (D).  Each reservoir is
connected to two edges.  We introduce a one dimensional coordinate
$x_\alpha$ that keeps track of the position along edge $\alpha$
($\alpha$ = i,ii,iii... in the example) growing continuously along the
direction of the arrows, i.e. from source to drain.  We denote by
$x_{\alpha i}$ the point where the lead coming from reservoir $i$
contacts the edge $\alpha$.  At each point along an edge we attach a
local $y$-axis normal to the edge.  The density varies rapidly as a
function of $y$ and slowly as a function of $x$.  We therefore
introduce an integrated {\it edge density fluctuation} \be
\label{integrateddensity}
\delta \hat \rho(x_\alpha)=\int dy~\delta \hat \rho (x_\alpha,y) \ee
where the edge is located at about $y=0$ and the
integral over $y$ extends far enough to include the whole region in
which the density fluctuations differ from zero (see
Fig. \ref{fig_probe.eps}).  From the algebra~(\ref{commutation2}) one
can then derive the commutation rules for edge density fluctuations.
In the Appendix \ref{app-rel} we show that \be \label{commutation}
[\delta \hat \rho(x_\alpha),\delta \hat \rho(x'_\beta)] =
-i\frac{\nu(x_\alpha) s_\alpha}{2\pi} \delta_{\alpha
\beta}\partial_{x_\alpha} \delta(x_\alpha-x'_\beta)~, \ee where
$s_\alpha=1$ if $\alpha$ is a left edge, $s_\alpha=-1$ if $\alpha$ is
a right edge, $\nu(x_\alpha)$ is the equilibrium filling factor
(defined as $\nu(x) = 2 \pi l^2 \rho_0(x)$) in the bulk contiguous to
the edge labelled ``$\alpha$".  Edge fluctuations on different edges
commute.

For constant $\nu$ Eq.~(\ref{commutation}) reproduces the Kac-Moody
current algebra for the density fields in the chiral Luttinger liquid
model.  Due to the presence of the noninteger factor $\nu$ these
commutation rules imply that each edge with $\nu<1$ exhibits $\chi$LL
behavior, even if all interactions between different edges are turned
off.  Notice that our derivation of the $\chi$LL has nothing to do
with the quantum Hall effect: it only depends on the coarse-graining
(the hydrodynamic approximation) and on the high magnetic field limit
(the projection in the LLL).  The derivation is valid for arbitrary
value of $\nu$, whereas the quantum Hall effect occurs only at special
values of $\nu$ for which, as will be argued below, the strength of
tunneling between different edge states becomes negligible.

In terms of edge density fluctuations, the hamiltonian becomes \be
H=\frac12\sum_{\alpha \beta}\int dx_\alpha \int dx'_\beta \delta \hat
\rho(x_\alpha) V(x_\alpha,x'_\beta)\delta \hat \rho(x'_\beta)
\label{hamiltonian-e}
\ee where $V(x_\alpha,x'_\beta) (=V(x'_\beta,x_\alpha))$ is
constructed from the original interaction $V(\rr,\rr')$ by putting
$\rr$ at position $x_\alpha$ along the edge $\alpha$, and $\rr'$ at
position $x'_\beta$ along the edge $\beta$.  By using
Eq.(\ref{commutation}) and the hamiltonian (\ref{hamiltonian-e}), the
equation of motion for edge density fluctuations is immediately found
to be \be \label{rhoequation} \partial_t \delta \hat \rho(x_\alpha) =
-\frac{\nu(x_\alpha) s_\alpha}{2\pi}\int_{-\infty}^\infty
dx'_\beta~\partial_{x_\alpha} V(x_\alpha,x'_\beta)\delta \hat \rho
(x'_\beta)~.  \ee Rather than pursuing the solution of Eq.~(\ref
{rhoequation}) in general, we shall henceforth restrict our attention
to the special case in which all the edges share the same bulk
density, i.e. $\nu(x_\alpha) = \nu$ independent of $\alpha$ and $x$.
It is convenient to define the ``displacement field" $\hat \phi
(x_\alpha,t)$ such that \be \delta \hat \rho(x_\alpha,t) = \partial_x
\hat \phi(x_\alpha,t).  \ee These fields satisfy the commutation
relations \be [\hat \phi(x_\alpha), \hat \phi(x'_\beta)] = i
\frac{\nu}{4 \pi} s_\alpha \sgn {x-x'}\delta_{\alpha \beta}~.  \ee
Assuming a time-dependence of the form \be \hat \phi(x_\alpha,t) =
\hat \phi(x_\alpha) e^{-i \omega t}~, \ee we see that
Eq.~(\ref{rhoequation}) takes the form \be i\omega \partial_{x_\alpha}
\hat \phi(x_\alpha) = \frac{\nu s_\alpha}{2\pi}
\int_{-\infty}^\infty dx_\beta' \partial_{x_\alpha}
V(x_\alpha,x'_\beta) \partial_{x_\beta'}\hat \phi(x'_\beta)~.
\label{eigenfun0}
\ee The associated eigenvalue problem \be i\omega
\partial_{x_\alpha}\varphi(x_\alpha) = \frac{\nu s_\alpha}{2\pi}
\int_{-\infty}^\infty dx_\beta' \partial_{x_\alpha}
V(x_\alpha,x'_\beta)\partial_{x'_\beta}\varphi(x'_\beta)~.
\label{eigenfun}
\ee is hermitian, and has the following properties:
\begin{enumerate}
\item All the eigenfrequencies $\omega_n$ are real.
\item If $\varphi_{n} (x_\alpha)$ is an eigenfunction with frequency
$\omega_n$ then $\varphi^*_{n} (x_\alpha)$ is an eigenfunction with
frequency $-\omega_n$.
\item The eigenfunctions $\varphi_{n\alpha} (x)$ form a complete basis
in the Hilbert space with the completeness \be -i\sum_n
\sgn{\omega_n}\varphi_{n}(x_\alpha)\varphi_{n}^*(x'_\beta)
\partial_{x'_\beta} = s_\alpha
\delta_{\alpha\beta}\delta(x_\alpha-x'_\beta).  \ee and the
orthonormality conditions \be \sum_\alpha is_\alpha\int dx_\alpha~
\varphi_{n}^*(x_\alpha) \partial_{x_\alpha} \varphi_{m}(x_\alpha) =
\sgn{\omega_n} \delta_{nm} \ee
\end{enumerate}
The proof of these relations is provided in the Appendix \ref{app-eig}
.

It is straightforward, with the help of the above relations, to show
that the density fluctuation field can be expanded on the basis
provided by the $\varphi_n$'s as follows \be \delta \hat
\rho(x_\alpha)=\sqrt{\frac{\nu}{2\pi}}\sum_{n>0}\left(\hat
b_{n}\partial_{x_\alpha}\varphi_{n}(x_\alpha)+ \hat b_{n}^\dagger
\partial_{x_\alpha}\varphi_{n}^{*}(x_\alpha)\right)~,
\label{rho-boson}
\ee where $n>0$ specifies that only the positive frequency
eigenfunctions are included in the sum and the $\hat b_n$s -- one for
each $n>0$ -- are boson operators obeying the standard commutation
relation $[\hat b_{n},\hat b_{n'}^\dagger ]= \delta_{nn'}$.  At the
same time, the hamiltonian~(\ref{hamiltonian-e}) takes the form \be
\hat H=\sum_{n>0}\hbar \omega_n \hat b_{n}^\dagger \hat b_{n}.  \ee

It is instructive at this point to solve the eigenvalue
equation~(\ref{eigenfun}) in a simple case.  We consider just two
parallel edges in a translationally invariant Hall bar geometry,
$\alpha =1$ for the left edge and $\alpha = 2$ for the right edge (see Fig. \ref{simple_probe.ps}).
The interaction is assumed to have the form \be V(x_\alpha-x'_\beta) =
\left(
\begin{array}{cc}
V_1(x_1-x'_1) & V_2(x_1-x'_2)\\ V_2(x_2-x'_1) & V_1(x_2-x'_2)
\end{array}
\right) \ee where $V_1$ and $V_2$ are translationally invariant
interactions between density fluctuations on the same edge and on
different edges respectively (The coordinates $x_1$, $x_2$ are set up
to have the same value on points at the same ``height" on the two
edges).  We seek the solutions of Eq.~(\ref{eigenfun}) in the form
$\varphi(x_\alpha) = \bar \varphi_\alpha (k) e^{ikx_\alpha}$.  This
leads to the $2 \times 2$ eigenvalue problem \be i \omega \left
(\begin{array}{cc} 1 & 0\\ 0& -1\end{array}\right) \left
(\begin{array}{c}\bar \varphi_1 \\ \bar \varphi_2 \end{array} \right)
= \frac{i k \nu}{2 \pi} \left(
\begin{array}{cc}
V_1(k) & V_2(k)\\ V_2(k) & V_1(k)
\end{array}
\right) \left (\begin{array}{c}\bar \varphi_1 \\ \bar \varphi_2
\end{array} \right)~, \ee where $V_1(k)$ and $V_2(k)$ are the Fourier
transforms of $V_1(x)$ and $V_2(x)$ (the upper part of the spinor
refers to the left edge).  The eigenvalues are \be \label{dispersion}
\omega_{k}=\pm \frac{\nu}{2\pi}\sqrt{V_1^2(k)-V_2^2(k)} \vert k \vert
~, \ee and for each positive frequency there are two solutions: the
``up-moving" one is \be \label{upwavefunction}
\varphi_k^u(x_\alpha)=\frac{1}{\sqrt{kL}}\left(
\begin{array}{c}
u_ke^{ikx_1} \\ -v_ke^{ikx_2}
\end{array}
\right) \ee with $k>0$, and the ``down-moving" one is \be
\label{downwavefunction}
\varphi_k^d(x_\alpha)=\frac{1}{\sqrt{kL}}\left(
\begin{array}{c}
v_ke^{-ikx_1} \\ -u_ke^{-ikx_2}
\end{array}
\right) ~, \ee also with $k>0$.  Here, as usual, we have normalized
the eigenfunctions with the factor $1/\sqrt{L}$ where $L$ is the
length of the edge. This length is assumed to be arbitrarily large,
and will not enter the physical results.  On the other hand the
presence of the normalization factor $1/\sqrt{k}$ is imposed by the
orthonormality conditions. Since these conditions require $u_k^2-v_k^2
=1$, one can write $u_k =\cosh \theta_k$, $v_k =\sinh \theta_k$ and
\be \tanh 2 \theta_k =\frac{V_2(k)}{V_1(k)}~.
\label{coherencefactors}
\ee Thus, we have recovered the standard expressions for the
dispersion of the edge waves in the ordinary (nonchiral) Luttinger
liquid model.  However, we emphasize that, in the present model, the
$\chi$LL behavior persists even if the interaction $V_2$ is turned
off.  This is due to the anomalous commutator~(\ref{commutation})
between density fluctuations on the same edge.

\section{Formulation of transport}
\label{transport-form}
In transport theory we need to calculate the change in the current
$I_i$ that flows into reservoir ``i" due to a change in the potential
$V_j$ of reservoir ``j" : \be \label{defconductance} \delta I_i =
\sum_j G_{ij} \delta V_j~.  \ee (The conductance matrix elements
$G_{ij}$ will in general depend on the initial values $V_i$ of the
applied voltages: these initial values will be referred to as ``bias
voltages".)  The current will be considered positive when it enters a
reservoir and negative when it leaves it.  Due to gauge invariance and
current conservation the $G_{ij}$s (under steady-state conditions)
satisfy the constraints \be \label{constraint} \sum_j G_{ij} =
\sum_iG_{ij} = 0~.  \ee These constraints specify the values of the
diagonal conductances $G_{ii}$ once the off-diagonal ones with $i \neq
j$ are known.
     
The form of the edge current-density operator $\hat I(x_\alpha)$ is
dictated by the continuity equation \be e\partial_t {\delta \hat
\rho}(x_\alpha)=\partial_x \hat I(x_\alpha)~, \ee which immediately
gives \be \hat I_{\alpha} (x) = e \partial_t \hat \phi_{\alpha}(x).
\ee This current density is positive when it flows along the direction
of the arrows in Fig. \ref{fig_probe.eps}, negative otherwise.  Thus,
the current flowing into terminal $i$ is given by \be
\label{terminalcurrent} \delta \hat I_i = \sum_\alpha \xi_{\alpha
i}\delta \hat I(x_{\alpha i}) \ee where \be \label{contactfunction}
\xi_{\alpha i}~=~ \left\{\begin{array}{c}~+1~~~
\mbox{if~$\alpha$~enters~$i$} \\ -1~~~\mbox{if~$\alpha$~exits~$i$} \\
~0~~~\mbox{otherwise}.
\end{array}\right.
\ee
\begin{widetext}

The linear response of the current density to a periodic variation of
the electrical potential $\delta V(x_\beta)$ is given by
\begin{eqnarray}
\label{linearresponse1}
\delta I(x_\alpha)~=~ i \frac{e^2}{\hbar} \sum_\beta \int
dx'_\beta~\delta V(x'_\beta) \int_0^\infty dt \langle
[\partial_t\hat \phi(x_\alpha,t),\partial_{x'_\beta}\hat
\phi(x'_\beta)]\rangle e^{i (\omega + i \eta)t} ~,
\end{eqnarray}
where $\omega$ is the frequency and $\langle...\rangle$ denotes the
equilibrium average.  A first integration by parts with respect to
time gives
\begin{eqnarray}  \label{linearresponse2}
\delta I(x_\alpha)~=~ i \frac{e^2}{h} \sum_\beta \int
dx'_\beta~[\omega \partial_{x'_\beta}D(x_\alpha,x'_\beta;\omega)]
\delta V(x'_\beta)~,
\end{eqnarray}
where
\begin{equation}\label{defD}
D(x_\alpha,x'_\beta;\omega) \equiv -2 \pi i \int_0^\infty
\langle [\hat \phi(x_\alpha,t),\hat \phi(x'_\beta)]\rangle
e^{i(\omega+i\eta)t}dt~
\end{equation}
 is the retarded displacement-field propagator, whose explicit
expression in terms of ``phonon eigenfunctions" is \be
\label{propagator}
D(x_\alpha,x'_\beta;\omega) ~=~ \nu \sum_{n>0} \left[
\frac{\varphi_{n}(x_\alpha)\varphi_{n}^{*}(x'_\beta)}{\omega-\omega_n+i\eta}
-
\frac{\varphi_{n}^{*}(x_\alpha)\varphi_{n}(x'_\beta)}{\omega+\omega_n+i\eta}
\right]~.  \ee

In doing the integral by parts we have exploited the fact that
$\partial_{x'}D(x_\alpha,x'_\beta;t=0^+)=$ $-2\pi i [\hat
\phi(x_\alpha),\partial_{x'_\beta}\hat \phi(x'_\beta)] \propto
\delta_{\alpha \beta}\delta(x_\alpha-x'_\beta)$ vanishes unless
$x_\alpha$ and $x'_\beta$ coincide.  It will be shown below that this
condition is always satisfied in the relevant region of integration.
\end{widetext}
A potential change $\delta V_j$ applied to the $j$-th lead can be
modelled as a change of the potential on the two edges that enter and
exit the reservoir.  The change in potential is considered uniform
over the portions of the edges that run inside the leads, and drops to
zero at the points of contact between the leads and the system (It
must be borne in mind that what we are modelling here is the
externally applied potential, not the full screened potential that
will appear all over the system in response to the external
perturbation).  Thus we see that the potential change associated with
reservoir $j$ is described by the equation \be
\label{terminalvoltage}
\partial_{x_\beta}\delta V(x_\beta) = \sum_j \xi_{\beta
j}\delta(x_\beta - x_{\beta j}) \delta V_j~, \ee where the ``contact
functions" $\xi_{\beta j}$ are defined in Eq.~(\ref{contactfunction}).

We now combine Eqs.~(\ref{linearresponse2}),~(\ref{terminalcurrent})
and~(\ref{terminalvoltage}).  The integral over $x'_\beta$ can be
immediately carried out (by parts) under the reasonable assumption
that the phonon eigenfunctions decay exponentially for $x \to \pm
\infty$ i.e., well inside the reservoirs.  This is physically expected
to happen as the one-dimensional edge channels broaden into a three
dimensional reservoir.  Mathematically, one must make sure that the
eigenfunctions used to calculate the displacement propagator satisfy
this boundary condition.  The final result for the current arriving at
reservoir $i$ via edge channel $\alpha$ due to a potential disturbance
applied to edge channel $\beta$ by reservoir $j$ with $i \neq j$ is
\be \delta I_i = \sum_j \left(-i \frac{e^2} {h} \sum_{\alpha
\beta}\xi_{\alpha i} \xi_{\beta j} \lim_{\omega \to 0}\omega D
(x_{\alpha i},x_{\beta j};\omega) \right) \delta V_j~.
\label{linearresponse3}
\ee The quantity within the round brackets is, by definition,
$G_{ij}$. Note that this equation specifies only the off-diagonal
elements ($i \neq j$) of the conductance matrix.  This guarantees that
$x_{\alpha i}$ is macroscopically distinct from $x_{\beta j}$ and
validates our integration by parts with respect to time.  Diagonal
elements $G_{ii}$ are determined by the continuity
conditions~(\ref{constraint}).

As a simple example consider the calculation of the propagator in the
translationally invariant geometry with short range interactions, so
that the phonon eigenfunctions are labelled by a wave vector $k$ and
$\omega_k = ck$ where $c$ is the velocity of the edge mode.  To ensure
that the phonon eigenfunctions vanish for $|x| \to \infty$ we shift
the wave vector $k$ infinitesimally into the complex plane, setting $k
\to k+i \eta {\rm sgn}(x)$ in Eq.~(\ref{upwavefunction}) and $k \to
k-i \eta {\rm sgn}(x)$ in Eq.~(\ref{downwavefunction}).
\begin{figure}[ht]
\begin{center}
\includegraphics[clip,height=6cm]{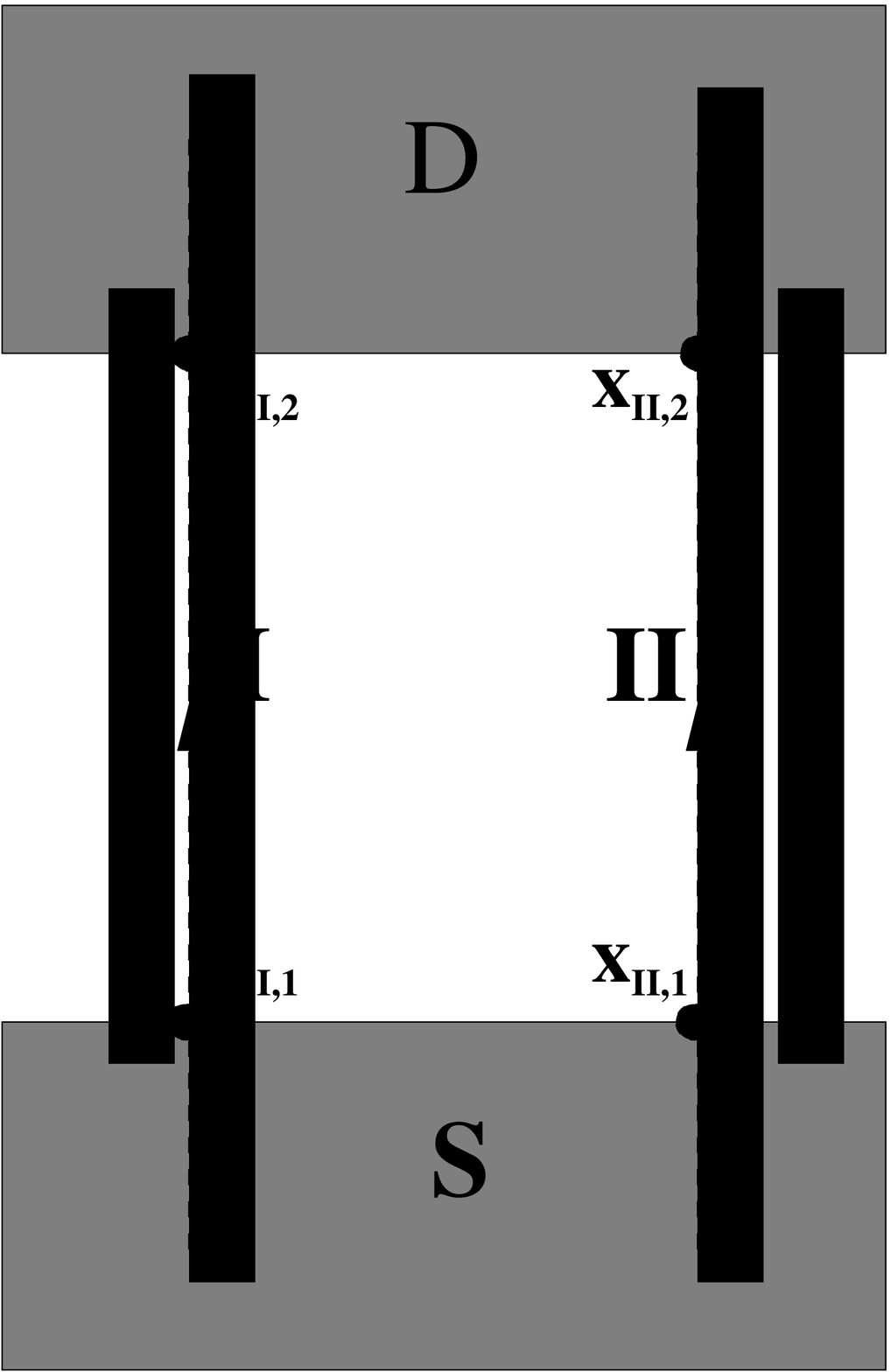}
\caption{Schematics of a translationally invariant two-terminal
device.}
\label{simple_probe.ps}
\end{center}
\end{figure}
Next, we substitute these eigenfunctions into Eq.~(\ref{propagator})
and convert the sum over $n$ into an integral along the real axis of
the complex variable $k$.  We readily find that only the poles at $k =
\pm(\omega + i\eta)/c$ contribute to the integral which thus yields
\begin{widetext}
\begin{eqnarray}
\label{simplepropagator}
- i\lim_{\omega \to 0} \omega D(x_\alpha,x'_\beta;\omega) ~=~ \nu
\Theta (x_\alpha-x'_\beta) \left ( \begin{array}{cc} u^2 & -uv \\-uv &
  v^2
\end{array} \right) + \nu \Theta (x'_\beta-x_\alpha) \left (
\begin{array}{cc} v^2 & -uv \\-uv & u^2 \end{array} \right)
\end{eqnarray}
where $u$, $v$ are the $k \to 0$ limits of $u_k$ and $v_k$.
\end{widetext}

Notice that in the case of decoupled edges ($u=1$, $v=0$) one has only
upward propagation on the left edge and downward propagation on the
right one.  This makes the conductance $G_{ij}$ vanish unless the
reservoirs $j$ is ``upstream" of reservoir $i$, consistent with the
definition of an {\it ideal quantum Hall system}\cite{Baranger1989}.
It is straightforward, at this point, to compute the two-terminal
conductances $G_{12}$, $G_{21}$ of the simple device shown in
Fig. \ref{simple_probe.ps}.  Since the source and the drain reservoirs
contact both edges, and $\xi_{\alpha 1}=-\xi_{\alpha 2}$ for each
edge, Eq.~(\ref{linearresponse3}) gives us \be \label{effectivenu}
G_{12} = \frac{e^2 \nu}{h} e^{-2 \theta} = G_{21}~.  \ee
Interestingly, the presence of the factor $e^{-2 \theta}=(u-v)^2$ in
the relation between the current and the source-drain potential does
not imply a deviation from the ideal Hall conductance, since the
relation between the Hall voltage (as measured by ideal voltage probes
applied to the two sides of the Hall bar) and the source-drain
potential is also modified by the same factor\cite{Wen1991b}.

\section{Reflection and transmission of edge waves}
\begin{figure}
\includegraphics[clip,width=7cm]{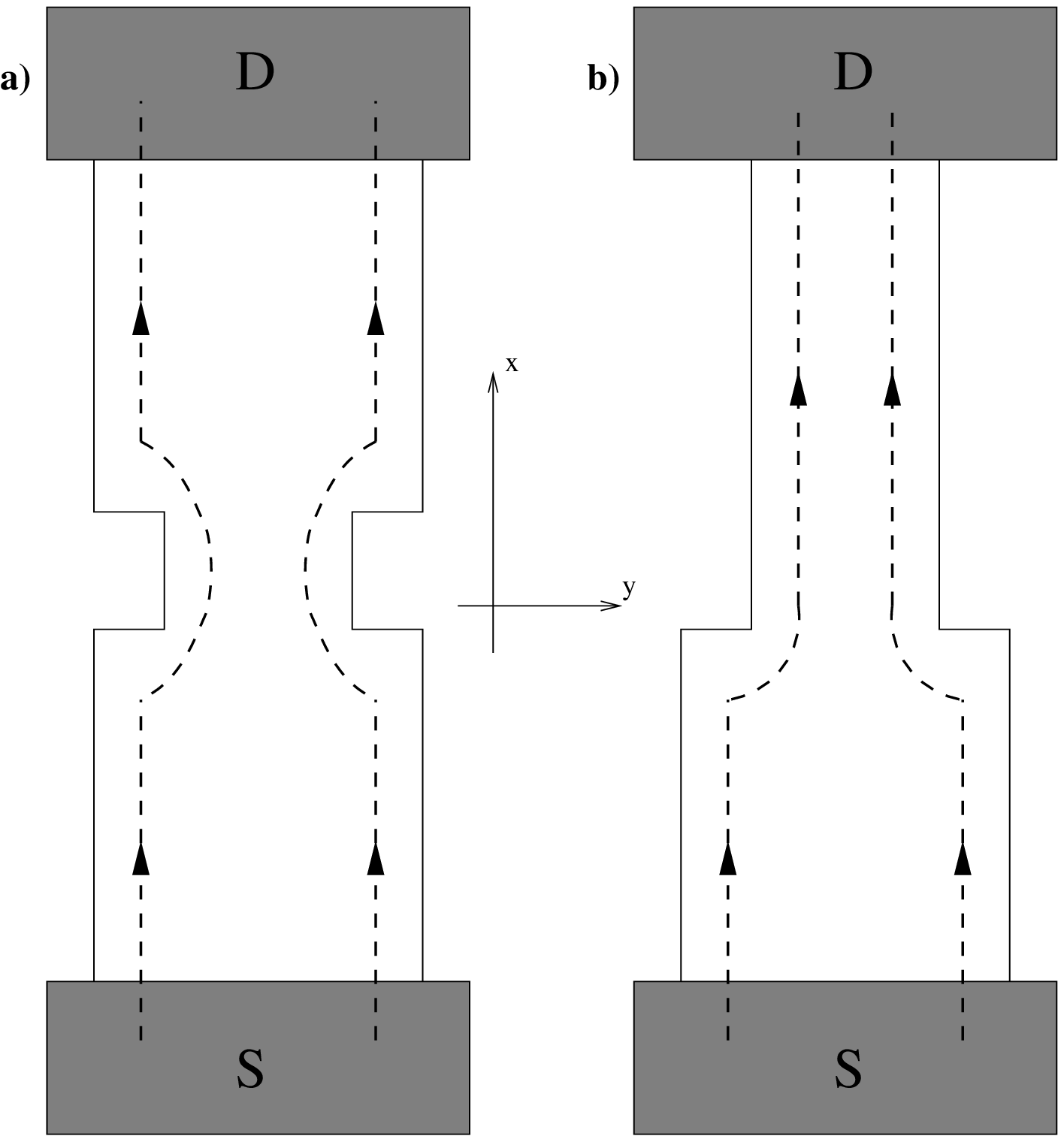}
\caption{The two types of constrictions we consider.  On the right, a
semi-infinite constriction. On the left, a more realistic constriction
localized in a finite region of the sample.  For clarity the lateral
voltage probes are not shown.}
\label{seminfinite}
\end{figure}

Before proceeding to the calculation of the conductances in the
presence of inter-edge tunneling we wish to take a closer look at free
edge waves in the presence of a constriction that breaks translational
symmetry (see Fig. \ref{seminfinite}).  This constriction can be
created by depleting a portion of the sample by applying a voltage to
metallic gates on top of the mesa. 
 When an edge wave of finite wave vector $k$ impinges on the
constriction it is partially reflected and partially transmitted. How
this affects the conductance depends crucially on the behavior of the
reflection coefficient $r(k)$ in the limit $k \to 0$.  If $r(0)=0$
then there is no correction to the ideal conductance (in the absence
of tunneling); otherwise there will be one.

To keep the analysis simple, we now assume that both $V_1$ and $V_2$
are short-ranged on the scale of the density variations: this means,
in particular, that only points at the same value of $x$ on opposite
edges interact and $V_{\alpha \beta}(x)$ has the form
\be V_{\alpha\beta}(x)= \left(
\begin{array}{cc}
V_1 & V_2(x)\\ V_2(x) & V_1
\end{array}
\right) \ee where \be V_2(x)=\left\{
\begin{array}{ll}
V_{2,1} & x<-d/2\\ V_{2,2} & |x|<d/2\\ V_{2,3}& x>d/2
\end{array}
\right. .  
\ee

Since the potential has a step-like behavior, with three different
values in the three regions $1$, $2$, and $3$, we seek the solution in
a piece-wise form. As in the standard scattering theory we label the
full solution with the quantum number of the incident wave. For
instance for an incident wave from the bottom on the left edge
$\varphi^u_{k_1}(x)$ we seek the ``up-moving" solution in the form \be
\label{scatteringup} \tilde\varphi_{k_1}^u(x)= \left\{
\begin{array}{ll}
\varphi_{k_1}^u(x)+r^u \varphi_{k_1}^d(x)& x<-d/2\\
A^u\varphi_{k_2}^{u}(x)+B^u \varphi_{k_2}^{d}(x)& |x|<d/2\\
t^{u}\varphi_{k_3}^{u}(x) & x>d/2
\end{array}
\right. \ee The wave vectors $k_1$, $k_2$, and $k_3$ in regions $1$,
$2$, and $3$ respectively are determined by the condition that the
energy of the wave is not changed in the scattering, i.e.  \be
\label{soundspeeds} c_1k_1=c_2k_2=c_3k_3 \ee where $c_1$, $c_2$, and
$c_3$ are the sound velocities in the corresponding regions. We remark
that the wave function $\tilde{\varphi}_{k_1}^u$ is labelled with the wave vector $k_1$ of the incident wave it 
originates from. In similar way one can construct the ``down-moving"
solution, 
\be \label{scatteringdown} \tilde\varphi_{k_3}^d(x)= \left\{
\begin{array}{ll}
t^d\varphi_{k_1}^d(x) & x<-d/2\\ A^d\varphi_{k_2}^{u}(x)+B^d
\varphi_{k_2}^{d}(x)& |x|<d/2\\ \varphi_{k_3}^{d}(x)+r^d
\varphi_{k_3}^{u}(x)& x>d/2
\end{array}
\right. , \ee
Notice that the spinor-like eigenfunctions
$\varphi_k^{u(d)}(x)$ (see Eqs.~(\ref{upwavefunction})
and~(\ref{downwavefunction})) are those appropriate for each region.
$r^{u(d)}$, $t^{u(d)}$ are the reflection and transmission amplitudes
for the up- (down-) moving wave. The matching conditions are dictated
by the physical requirement that there is no accumulation of energy at
the interfaces. This is equivalent to the requirement of continuity of
the solution at $x=\pm d/2$ and gives {\it four} conditions from which
the coefficients $A$, $B$, $t$ and $r$ can be determined. The
solution, expressed in terms of the mixing angles, is
\begin{widetext}
\be
\begin{split}
t^u
&=\frac{e^{-i\frac{k_1+k_3}{2}d}}{\cos(k_2d)\cosh(\theta_1-\theta_3)-i
\sin(k_2d)\cosh(2\theta_2-\theta_1-\theta_3)}\sqrt{\frac{c_1}{c_3}},\\
r^u &=-\frac{\cos(k_2d)\sinh(\theta_1-\theta_3)+i \sin(k_2
d)\sinh(2\theta_2-\theta_1-\theta_3)}
{\cos(k_2d)\cosh(\theta_1-\theta_3)-i
\sin(k_2d)\cosh(2\theta_2-\theta_1-\theta_3)}e^{-ik_1d},\\ A^u
&=\frac{\cosh(\theta_3-\theta_2)e^{-i\frac{k_1+k_2}{2}d}}
{\cos(k_2d)\cosh(\theta_1-\theta_3)-i
\sin(k_2d)\cosh(2\theta_2-\theta_1-\theta_3)}\sqrt{\frac{c_1}{c_2}},\\
B^u &=\frac{\sinh(\theta_3-\theta_2)e^{i\frac{k_2-k_1}{2}d}}
{\cos(k_2d)\cosh(\theta_1-\theta_3)-i
\sin(k_2d)\cosh(2\theta_2-\theta_1-\theta_3)}\sqrt{\frac{c_1}{c_2}}.
\label{solution}
\end{split}
\ee
\end{widetext}
The expression for the coefficients in the ``down-moving" solution can
be obtained with the substitutions
\begin{equation}
t^u\rightarrow t^d;~r^u\rightarrow r^d;~ A^u\rightarrow
B^d;~B^u\rightarrow A^d
\end{equation}
and \be \theta_1\to \theta_3,~\theta_3\to\theta_1,~c_1 \to c_3.
\label{coeff-transf}
\ee The reflection and transmission coefficients satisfy the relation
$|r|^2+(c_1/c_3)|t|^2=1$ which follows from the conservation law
derived in Appendix \ref{con_law}.  By sending $d$ to zero one can
examine the case described in the left panel of
Fig. \ref{seminfinite}. When $d=0$, one gets \cite{Oreg1995}
\be
\begin{split}
t^u &=\frac{1}{\cosh(\theta_1-\theta_3)}\sqrt{\frac{c_1}{c_3}}\\ r^u
&=-\frac{\sinh(\theta_1-\theta_3)}{\cosh(\theta_1-\theta_3)}.
\label{rtsemiinf}
\end{split}
\ee In Eq.(\ref{effectivenu}) we have seen that the inter-edge
interaction renormalizes the two terminal conductance $G_{12}$ with
the factor $e^{-2\theta}$ which gives rise to an effective filling
factor $\tilde\nu=\nu e^{-2\theta}$. From this point of view by
introducing the effective filling factor in the different regions
$\tilde{\nu}_i= \nu e^{-2\theta_i}~(i=1,2,3)$, one can rewrite
(\ref{rtsemiinf}) as \cite{Chklovskii1998} \be
\begin{split}
r&=\frac{\tilde\nu_1-\tilde\nu_3}{\tilde\nu_1+\tilde\nu_3},\\
t&=\frac{2\tilde\nu_3}{\tilde\nu_1+\tilde\nu_3}\sqrt{\frac{c_1}{c_3}}.
\end{split}
\ee

We can now ask how the reflection of edge waves modifies the
conductances obtained for the translationally invariant case at the
end of the previous section.  To keep the discussion as simple as
possible consider first the situation in which the inter-edge
interaction is present only in the constriction region
$-d/2<x<d/2$. Four reservoirs are attached to the system above and
below the constriction.  A straightforward calculation
with eigenfunctions of the form~(\ref{scatteringup}) and~(\ref
{scatteringdown}) gives the following expressions for the
dimensionless conductances $g_{ij}=\frac{hG_{ij}}{\nu e^2}$: \be
\begin{split}
g_{21}&=\frac{1}{2\pi i}\int_{-\infty}^{\infty}dk
\frac{e^{ik(x_2-x_1)}}{k-\omega /c -{\rm i}0^+} t^u (k)\\ g_{31}&=0\\
g_{41}&=\frac{1}{2\pi i }\int_{-\infty}^{\infty}dk
\frac{e^{-{\rm i}k(x_4+x_1)}}{k-\omega /c -{\rm i}0^+}r^u (k)
\end{split}
\label{conductancescostriction} \ee 
with the transmission and reflection coefficients given by
Eq.(\ref{solution}) evaluated for $\theta_1=\theta_3=0$. They read \be
\begin{split}
t^u (k) &=\frac{{\rm i}e^{ikd}} {{\rm
i}\cos(c_1kd/c_2)+\sin(c_1kd/c_2)\cosh(2\theta_2)} \\ r^u(k)
&=\frac{\sin(c_1kd/c_2)\sinh(2\theta_2)e^{-{\rm i}kd}} {{\rm
i}\cos(c_1kd/c_2)+\sin(c_1kd/c_2)\cosh(2\theta_2)}.
\end{split}
\label{tr} \ee 
The key observation at this point is that the
exponential factors in Eq.(\ref{conductancescostriction}) force to
close the integration contours in the upper half-plane. Apart from the
pole at $k=\omega /c +{\rm i}0^+$ there are no other poles in the
upper half-plane since both $t^u(k)$ and $r^u(k)$ have poles in the
lower half-plane. Thus, we obtain 
\be
\begin{split}
G_{21}&= \frac{e^2 \nu}{h}t^u(\omega /c) \\ G_{41}&= \frac{e^2
\nu}{h}r^u(\omega /c)
\end{split}
\label{gsconstriction} 
\ee 
and in the limit of zero frequency one
recovers the exact quantization that characterizes the ideal
fractional QHE.  This result can be understood by observing that in
the long wavelength limit the constriction becomes fully transparent
to the current.  The situation is quite different in the case of the
semi-infinite constriction where $r(0)$ acquires a finite value: in
this case one finds deviations from the ideal Hall conductance.  It is
amusing to see that the expressions~(\ref{gsconstriction}) are similar
in form to what the Landauer-B\"uttiker theory would predict for a
situation in which particles are physically backscattered from one
edge to the other with probability $r(0)$.  However, up to this point,
our theory does not allow for the transfer of charge between the
edges.

\section{The tunneling hamiltonian}
It is now time to consider the effect of charge tunneling between
different edges.  The physical origin of tunneling lies in the fact
that the electron quasiparticles are not $100\%$ localized on one or
the other edge: the density matrix $\rho(\rr,\rr') = \langle \hat
\Psi^\dagger(\rr) \hat \Psi(\rr')\rangle$ ($ \hat \Psi^\dagger(\rr)$
is the creation operator of a quasiparticle at position $\rr$) has a
finite value even when $\rr$ and $\rr'$ are on different edges. This
is true at all filling factors but, of course, the range of the
density matrix depends dramatically on whether there are extended
quasiparticle states at the chemical potential, and this in turn
depends on the filling factor.  The fractional QHE is believed to
arise at electronic densities such that there are no extended
quasiparticle states at the chemical potential, so that
$\rho(\rr,\rr')$ is exponentially small when $\rr$ and $\rr'$ (on
different edges) are separated by a distance larger than $\sim \ell$.
This means that there is essentially no tunneling between the edges.
Even in this case, however, tunneling can be induced by pushing two
edges together as in the constrictions studied in the previous
section.

Since the physics of tunneling is lost in the hydrodynamic
approximation, (which is local in space and therefore does not allow
for any direct connection between the edges) we need to put it back in
the hamiltonian ``by hand". To this end we define a quasiparticle
operator $\hat \Psi^\dagger_\alpha(x_\alpha)$ which adds a charge
$e^*$ (not necessarily equal to the electron charge $e$) localized at
position $x_\alpha$ along the $\alpha$ edge. This is accomplished by
requiring that $\hat \Psi^\dagger_\alpha(x_\alpha)$ satisfy the
commutation relation \be
\left[\Psi_\alpha^\dagger(x_\alpha),\delta\rho(x'_\beta)\right]=-\frac{e^*}{e}
\delta_{\alpha\beta}\delta(x_\alpha-x'_\beta)
\Psi_\alpha^\dagger(x_\alpha).
\label{psi-com-rho} \ee We hasten to say that we do not know, in
general, what the correct value of $e^*$ is.  In some special cases,
for example at filling factors of the form $\nu = \frac{1}{2n+1}$ with
integer $n$, it is widely believed that $e^*=\nu e$, but there is no
general theory for arbitrary filling factors.  Let us then treat $e^*$
as a phenomenological parameter, and note that Eq.~(\ref{psi-com-rho})
is satisfied by\footnote{Notice that we have chosen the filling factor 
constant and
we have written this operator already in a normal ordered form. We have not explicitely written a normalization factor depending on a short-distance cut-off.
%This
%is useful in avoiding some complications with the
%normalization.
}
%{ Indeed without taking into account the normal ordering
%an infrared divergence appears and this requires a new definition of
%the correlation function \cite{Luther1974}.}
\be
\begin{split}\label{psisol}
\Psi_\alpha^\dagger(x_\alpha)=&\hat U^\dagger_\alpha
\exp\left[-i\frac{e^*}{e} \sqrt{\frac{2\pi}{\nu}}s_{\alpha}
\sum_{n>0}\varphi_{n}^{*}(x_\alpha)\hat b_{n}^\dagger\right]\\ &\times
\exp\left[-i\frac{e^*}{e}\sqrt{\frac{2\pi}{\nu}}s_{\alpha}
\sum_{n>0}\varphi_{n}(x_\alpha)\hat b_{n}\right]~
\end{split}
\ee where $\hat U^\dagger_\alpha$ is an operator that commutes with
all the $\hat b$'s and $\hat b^\dagger$'s and increases the total
charge $\hat Q_\alpha$ on the edge by $-e^*$
\be [\hat
U^\dagger_\alpha, \hat Q_\beta] = {e^*}\delta_{\alpha \beta} \hat
U^\dagger_\alpha~.  \ee The statistics of the quasiparticle is
determined by the charge. If $e^*=e$ and $\nu=1$ the fermion
commutation relations are satisfied, otherwise the quasiparticle has a
fractional charge and a fractional statistics.  Another point to be
made is that the creation of a fractional charge is not in
contradiction with the quantization of the electric charge: the fact
that the charge on the edge varies by a fractional amount simply means
that a compensating fractional variation must be occurring deep in the
reservoirs to keep the total charge in the universe an integer.

In terms of the quasiparticle operators, the tunneling between two
edges (``left" and ``right") coupled by a constriction at $x=0$ is
described by the tunneling hamiltonian \be \label{htun} \hat H_T=
\Gamma :\hat\Psi^\dagger _L(0)\hat \Psi_R(0): +\Gamma^* :\hat
\Psi^\dagger_R(0) \hat \Psi_L(0):, \ee where $\Gamma$ is a
(phenomenological) tunneling amplitude and $:\ldots:$ indicates the
normal ordering.  One can of course consider more general situations
in which tunneling occurs simultaneously at different points.

\section{The tunneling conductance}
Let us consider the case of just two edges coupled by a constriction
at $x=0$.  The complete hamiltonian is \be \hat H = \sum_{n>0}\hbar
\omega_n \hat b^\dagger_n \hat b_n + \hat H_T~, \ee where the
tunneling term $\hat H_T$, given by Eq.~(\ref{htun}), introduces an
interaction between the formerly free bosons $\hat b_n$. At the same
time, the total charge on, say, the left edge is no longer a constant
of motion: its time derivative defines the {\it tunneling current}
$\hat I_T$ as follows:
\begin{eqnarray}
\hat I_T &=& -\frac{i}{\hbar} \left[\hat Q_L, \hat H_T\right]\nonumber
\\ &=&i \frac{e^*}{\hbar}\left(\Gamma \hat \Psi_L^\dagger(0)\hat
\Psi_R(0)- \Gamma^* \hat \Psi_R^\dagger(0)\hat \Psi_L(0)\right).
\end{eqnarray}

Looking back at Eq.~(\ref{linearresponse3}) we see that the task at
hand is that of calculating the correction to the displacement field
propagator due to the interaction between the bosons.  We will now
show that this correction can be exactly expressed in terms of a
tunneling current propagator and will provide a perturbative
evaluation of the latter.

First, let us introduce some compact notation.  We define \be \hat
B_n^i \equiv \left( \begin{array}{c} \hat b_n\\ \hat
b^\dagger_n\end{array}\right)~, \ee where $i = 1(2)$ for the upper
(lower) component, and the associated phonon propagator \be {\cal
D}^{ij}_{nn'}(t) \equiv -\frac{i}{\hbar} \Theta (t)\langle[\hat
B_n^i(t),\hat B_{n'}^{\dagger j}]\rangle~ .  \ee Similarly we define
\be \varphi_{n}^i (x_\alpha) \equiv \left( \begin{array}{c}
\varphi_{n}(x_\alpha)\\ \varphi^*_{n}(x_\alpha)\end{array}\right)~,
\ee so that the phonon field propagator can be written as \be
\label{phipropagator} D(x_\alpha,x_\beta',t) = \hbar\nu
\varphi_{n}^i (x_\alpha) {\cal D}^{ij}_{nn'}(t)\varphi_{n}^j
(x_\beta') \ee (sum over repeated indices).

The phonon operators satisfy the equation of motion \be i \partial_t
\hat B^i_n = \Omega_n^{ij}\hat B^j_n -\frac{Y^i_n}{e} \hat I_T~, \ee
where \be \hat \Omega_n^{ij}= \left(\begin{array}{cc}
\omega_n&0\\0&-\omega_n \end{array}\right)~,~~~~Y^i_n = \left(
\begin{array}{c} ~\gamma_n\\-\gamma^*_n \end{array}\right)~, \ee
and \be \gamma_n = \sqrt{\frac{2 \pi}{\nu}}\sum_\alpha
\varphi^*_{n}(0_\alpha)~.\ee

Then it is straightforward to verify that the phonon propagator
satisfies the equation of motion \be \left(i \partial_t \delta_{il} -
\Omega_n^{il} \right){\cal D}_{nn'}^{lj} =
\frac{(-1)^i}{\hbar}\delta_{ij}\delta_{nn'} \delta(t) -\frac{Y^i_n}{e}
{\cal G}^j_{n'}(t) \ee where the auxiliary propagator \be {\cal
G}^j_{n}(t)= -\frac{i}{\hbar} \Theta(t) \langle[\hat I_T(t),\hat
B^{\dagger j}_n]\rangle \ee satisfies in turn the equation of motion
\be \left(i \partial_t \delta_{ij} - \Omega_n^{ij} \right){\cal
G}^j_{n}(t) = - \frac{(Y^i_n)^*}{e}\tilde M_T(t) \ee with \be \tilde
M_T(t) = M_T(t)-\frac{{e^*}^2}{\hbar^2}\langle \hat H_T\rangle
\delta(t)
\label{mttunneling}
\ee and \be M_T(t) = - \frac{i}{\hbar} \Theta(t) \langle [\hat
I_T(t),\hat I_T]\rangle~ \ee is the {\it tunneling current
propagator}.

This system of equations is readily solved by Fourier transformation
with the following result
\begin{equation}\label{interactingphononpropagator}
\begin{split}
{\cal D}_{nn'}^{ij}(\omega) =& [{\cal
D}^{(0)}]_{nn'}^{ij}(\omega)+\frac{\hbar^2}{e^2} [{\cal
D}^{(0)}]_{nn_1}^{il}(\omega)\\ &\times Y^l_{n_1}\tilde M_T(\omega)
(Y^m_{n_2})^* \left([{\cal D}^{(0)}]_{n_2n'}^{mj}(\omega)\right)^*~,
\end{split}
\ee where $ [{\cal D}^{(0)}]_{nn'}^{ij}(\omega)$ is the noninteracting
phonon propagator.  Thus the tunneling correction to the phonon
propagator is expressed in terms of the tunneling current propagator,
as promised.

We can now make use of this result to calculate the {\it correction}
to the ideal conductances obtained in section III.  Let us denote by
$G^{(0)}_{ij}$ the conductance obtained in the absence of tunneling
and by \be \label{deltaGdef} \delta G_{ij}=G_{ij} - G^{(0)}_{ij} \ee
the correction due to the tunneling.  After some straightforward
manipulations one arrives at
\begin{equation}\label{deltaG}
\begin{split}
\delta G_{ij} =& -\frac {i}{\nu^2} \lim_{\omega \to 0} \sum_{\alpha_i
\gamma}\xi_{\alpha i} [D^{(0)}](x_{\alpha i},0_\gamma;\omega)\\
&\times [\omega\tilde M_T(\omega)]\sum_{ \delta
\beta_j}[D^{(0)}]^*(0_\delta,x_{\beta j}';\omega)\xi_{\beta j}
\end{split}
\ee where the indices $\gamma$ and $\delta$ run over the two edges
that are coupled by tunneling at $x=0$, and the Green's function of
the noninteracting displacement field, $[D^{(0)}]_{\alpha
\beta}(x,x';\omega)$, is given by Eq.~(\ref{propagator}).

As a concrete example, consider a four-terminal geometry, as may be
obtained from Fig. \ref{fig_probe.eps} by considering only terminals
1-4.  Assume for simplicity that the mixing angle $\theta$ is
independent of $x$. Then from Eq.~(\ref{simplepropagator}) we
immediately get
\begin{equation}
\begin{split}
\sum_{\gamma} [D^{(0)}](x_\alpha,0_\gamma;\omega) =& \sum_{\gamma}
[D^{(0)}](0_\gamma,-x_\alpha;\omega)\\ =& \frac{i
\nu}{\omega}e^{-\theta}\left[\Theta(x)
\left(\begin{array}{c}~u\\-v\end{array}\right)\right.\\ &\left.+
\Theta(-x) \left(\begin{array}{c}-v\\ ~u\end{array}\right)\right]
\end{split}
\ee where the upper (lower) component refers to the left (right) edge
and $u=\cosh\theta$, $v=\sinh\theta$.
\begin{widetext}
Substituting this in Eq.~(\ref{deltaG}) we find \be \delta
G_{ij}\label{linearresponse4} = \sum_{\alpha_i \beta_j} \delta
G_{\alpha_i \beta_j}(x_i,x_j)\xi_{\alpha i}\xi_{\beta j}~, \ee where
\begin{equation}\label{deltaG4}
\begin{split}
\delta G_{\alpha \beta}(x,x') =& -i e^{-2 \theta} \lim_{\omega \to
0}\frac{\tilde M_T(\omega)}{\omega} \left\{ \Theta(x)\Theta(-x')
\left(\begin{array}{cc}u^2~&-uv\\-uv&v^2~\end{array} \right)+
\Theta(-x)\Theta(x')
\left(\begin{array}{cc}~v^2&-uv\\-uv&~u^2\end{array} \right)\right.\\
&\left. + \Theta(x)\Theta(x') \left(\begin{array}{cc}-uv&u^2~\\
~v^2&-uv\end{array} \right)+ \Theta(-x)\Theta(-x')
\left(\begin{array}{cc}-uv&~v^2\\ u^2~&-uv\end{array}
\right)\right\}.
\end{split}
\ee Putting this in Eq.~(\ref{linearresponse4}) and noting that
$\xi_{\alpha 1}=\xi_{\beta 4}=1$, $\xi_{\alpha 2}=\xi_{\beta 3}=-1$
(with the labels $i$, $j$ as specified in the figure) we finally
obtain the correction to the Landauer-B\"uttiker conductances of the
ideal system: \be \label{deltaGideal} \delta G_{ij}= i
e^{-2\theta}\lim_{\omega \to 0} \frac{{\tilde M}_T(\omega)}{\omega}
\left(\begin{array}{cccc} uv&v^2&-uv&-v^2\\
u^2&uv&-u^2&-uv\\-uv&-v^2&uv&v^2\\ -u^2&-uv&u^2&uv
\end{array}\right)~. \ee

In Appendix D we show that \be g_T\equiv i\lim_{\omega\to
0}\frac{\tilde M_T(\omega)}{\omega}= 4\frac{|\Gamma|^2
{e^*}^2}{\hbar^3}\int_0^\infty dt~ t \mathrm{Im}G_-(t)~,
\label{gtunneling}
\ee where $G_-(t)=G_-(0,t;0,0)$ and \be
%\begin{split}
G_-(x,t;x',t')= \left\langle :\Psi_L^\dagger(x',t') \Psi_R(x',t'):
:\Psi_R^\dagger(x,t)\Psi_L(x,t):\right\rangle .
\label{correlator}
%\end{split}
\ee Thus, the complete set of conductances has been expressed in terms
of equilibrium averages of quasiparticle operators.

Notice that the presence of a bias voltage $V_\alpha$ on the edge
$\alpha$ modifies the time evolution of the corresponding
quasiparticle operator from $\hat \Psi^\dagger_\alpha (x_\alpha,t)$ to
$\hat \Psi^\dagger_\alpha (x_\alpha,t)e^{- i\frac{e^* V_\alpha
t}{\hbar}}$.  The underlying physical assumption is, of course, that
each edge is in equilibrium with a reservoir at potential $V_\alpha$.
Under this assumption, the bias voltage dependence of the conductances
can be calculated with no additional effort.

To conclude this section we consider a specific experimental setup of
Fig. 1.\cite{Roddaro2002}
%{\bf Roberto we must modify Fig. 1 to include the constriction}.  
The resistance $R_{xx}$ of the quantum point contact is measured
between terminals 3 and 4 in Fig. \ref{fig_probe.eps} \be
R_{xx}=\frac{V_4-V_3}{I}, \ee where $I$ is the source-to-drain
current.  By considering that the constriction does not affect the
source and drain probes, the full conductance matrix reads \be
\label{gij} G_{ij}=\frac{e^2}{h}\nu\left(
\begin{array}{cccccc}
1 & 0 & 0 & 0 & 0 & -1\\ 0 & 1 & 0 & -1 & 0 & 0\\ -1 & 0 & 1+\delta
g_{11} & \delta g_{12} & \delta g_{13} & \delta g_{14}\\ 0 & 0 &
-1+\delta g_{21} & 1+\delta g_{22} & \delta g_{23} & \delta g_{24}\\ 0
& -1 & \delta g_{31} & \delta g_{32} & 1+\delta g_{33} & \delta
g_{34}\\ 0 & 0 & \delta g_{41} & \delta g_{42} & -1+\delta g_{43} &
1+\delta g_{44}
\end{array}
\right),  
\ee 
where the indices $i$, $j$ run over $\{S,D,1,2,3,4\}$
and the right bottom submatrix is given by $\delta g_{ij} = \frac{h
\delta G_{ij}}{\nu e^2}$.  As it is customary in the experimental
setup we fix $V_D=0$, $I_S=-I_D=-I$ and $I_1=I_2=I_3=I_4=0$.
%{\bf Roberto,  note the changed sign of $I$, consistent with our conventions. 
% This might solve your sign problem}.   
With these constraints, the equation~(\ref{defconductance}) can be
easily solved, and to the lowest non vanishing order in $g_T$ we get
\be R_{xx}=\frac{h^2}{\nu^2 e^4} e^{-2\theta}(u+v)u~
g_T=\frac{h^2}{\nu^2 e^4} e^{-\theta}\cosh{\theta}~ g_T.
\label{tunnel-resistor}
\ee Notice that this perturbative result is valid only so long as
$R_{xx}$ is much smaller that $\frac{h}{e^2}$: the tunneling amplitude
$\Gamma$ must be sufficiently small for this to happen.

\section{Tunneling in the presence of a constriction}
Let us apply the formalism developed in the previous section to
evaluate the resistance of a constriction of the type shown in
Fig.~\ref{fig_probe.eps}. The mixing angles $\theta_1$ and
$\theta_3$ in the two external regions are assumed to be equal, while 
$\theta_2$ ($\theta_2>\theta_1$) measures the strength
of the interaction within the constriction.

The calculation of the correlation function $G_\pm$ can be perfomed by
using the definition (\ref{psisol}) for $\Psi^\dagger$ and the
Haussdorf lemma \be e^A e^B=e^B e^A e^{[A,B]}.  \ee We start by
considering the zero temperature limit where we obtain \be
\begin{split}
G_-(x,t;x',t')= \exp\left[\frac{2\pi{e^*}^2}{\nu e^2}\sum_{\lambda,
k>0}\left(
\tilde\varphi_{kR}^\lambda(x)+\tilde\varphi_{kL}^\lambda(x)\right)
\left(\tilde\varphi_{kR}^{\lambda*}(x')+\tilde\varphi_{kL}^{\lambda*}(x')\right)
e^{i\omega_k(t-t')} \right].
\end{split}
\ee When we substitute the functions $\tilde\varphi_k^{u(d)}(x)$ with
those determined in eqs. (\ref {scatteringup}) and
(\ref{scatteringdown}) we obtain (we use $x=x'$ with $x$ inside the
region of the constriction) \be
\begin{split}
G_-(x,t;x,t')=&\exp \left[\frac{2\pi{e^*}^2}{L\nu e^2}e^{-2\theta_2}
\sum_{k_2>0}\frac{1}{k_2} \left(|A^u e^{ik_2x}-B^ue^{-ik_2x}|^2+|A^d
e^{-ik_2x}-B^de^{ik_2x}|^2\right) e^{ik_2 c_2 (t-t')}\right]
\end{split}
\ee where the coefficients $A^u$, $B^u$, $A^d$ and $B^d$ are given by
(\ref{solution})-(\ref{coeff-transf}).  By assuming that the tunneling
is localized only at the point $x=x'=0$ and substituting the
expression (\ref{solution}) in this equation we obtain the key result
\be \label{finalG}
\begin{split}
G_-(t)=&\exp \left[\frac{4\pi{e^*}^2}{L\nu
e^2}\frac{\cosh(2\theta_2)}{\cosh(2\theta_1)}
e^{-2\theta_2}\sum_{k_2>0}
\left(\frac{\cosh(2\theta_{12})-\sinh(2\theta_{12})\cos(k_2d)}
{1+2\sinh^2(\theta_{12})\sin^2(k_2d)}\right)\frac{e^{ik_2 c_2 t}}{k_2}
\right]
\end{split}  \ee where we have defined
$\theta_{12}=\theta_1-\theta_2$.  For the function $G_+(t)$ the
calculation is similar and we can obtain $G_+(t)$ from the above
expression with the substitution $t\to -t$.
\end{widetext}
Before going into the detailed analysis of the above expression, it is
useful to recall that in the limiting case $\theta_{12}=0$ we recover
the result of Wen \cite{Wen1991b} for the case of interacting edges
\be
\begin{split}
G_-^W(t)&=\exp \left[\frac{4\pi {e^*}^2}{L\nu
e^2}e^{-2\theta_1}\sum_{k>0}\frac{e^{i\omega_{k}t}}{k}  \right].
\end{split}
\label{gofwen} 
\ee Notice that the presence of the inter-edge interaction leads to a
renormalization of the power-law
behavior of the current-voltage characteristics via the factor
$e^{-2\theta_1}$.  The explicit form of the function $G^W_\pm(t)$ can
be obtained by using the well known analytical results
\begin{equation}
\begin{split}
&\sum_{n=1}^\infty \frac{\cos(n q)}{n}=-\frac12 \ln(2-2\cos(q)),\\
&\sum_{n=1}^\infty \frac{\sin(n q)}{n}=\frac12 (\pi-q).
\end{split}
\end{equation}
If $q$ is a small quantity we have the approximate results
\begin{equation}
\begin{split}
\sum_{n=1}^\infty \frac{\cos(n q)}{n}\simeq-\ln(q), ~\sum_{n=1}^\infty
\frac{\sin(n q)}{n}\simeq\frac{\pi}2.
\end{split}
\end{equation}
In the case of the expression (\ref{gofwen}) we can evaluate the
series, after defining the integer $j$ as $j=k_1L/2\pi$, obtaining \be
G_\pm^W(t)=\exp\left[-\frac{2{e^*}^2}{\nu
e^2}e^{-2\theta_1}\ln\left(1-e^{\mp\frac{2\pi i c_1
t}{L}-\delta}\right)\right] \ee and in the limit of large system size
$ct/L\ll 1$ we have \be G_\pm^W(t)=\left(\delta\pm\frac{2\pi
ict}{L}\right)^{-\frac{2}{\nu}\frac{{e^*}^2}{e^2}e^{-2\theta_1}} \ee
where $\delta$ assures the convergence of the series even when $t\to
0$. This function is the propagator for the Luttinger Liquid model,
with the anomalous exponent
$\frac{2}{\nu}\frac{{e^*}^2}{e^2}e^{-2\theta_1}$. Notice that if we
assume $e^*=\nu e$ we get for this exponent $2\nu
e^{-2\theta_1}=2\tilde\nu$.  In this case the tunneling differential
conductance at zero temperature is predicted to have a power law
behavior with exponent given by $2({e^*/e})^2/\nu-2$.  Let us go back
to Eq.(\ref{finalG}). First we notice that the additional
$k$-dependent factor in the sum of Eq.(\ref{finalG}) does not alter
the logarithmic behavior at long times. To see this we define the
quantity
\begin{equation}
\begin{split}
S_-(t,d)=&\frac{2\pi}{L}\frac{\cosh(2\theta_2)}{\cosh(2\theta_1)}\\
&\times \sum_{k_2>0}\frac{e^{ik_2 c_2t}}{k_2}\\
&\times\left(\frac{\cosh(2\theta_{12})-\sinh(2\theta_{12}) \cos(k_2d)}
{1+2\sinh^2(\theta_{12})\sin^2(k_2d)}\right)
\end{split}
\end{equation}
and evaluate it numerically. To do this we calculate separately its
real and imaginary part,
\begin{equation}
\begin{split}
ReS_-(t,d)=&\sum_{n=1}^\infty\frac{\cosh(2\theta_{12})-\sinh(2\theta_{12})
\cos(2\pi dn/L_2)}{1+2\sinh^2\theta_{12}\sin^2(2\pi dn/L_2)}\\
&\times\frac{\cos(2\pi n c_2t/L)}{n},\\
ImS_-(t,d)=&\sum_{n=1}^\infty\frac{\cosh(2\theta_{12})-\sinh(2\theta_{12})\cos(
2\pi dn/L_2)}{1+2\sinh^2\theta_{12}\sin^2(2\pi dn/L_2)}\\ &\times
\frac{\sin(2\pi c_2tn/L)}{n}. \label{sum}
\end{split}
\end{equation}
Notice that in these sums we have substituted $k_2=2\pi n/L_2$ where
$n$ is an integer 
and \be \label{defL2} {L_2}=L
\frac{\cosh(2\theta_1)}{\cosh(2\theta_2)} \ee takes into account the
different speed of propagation of the waves in regions 1 and 2 (see
Eq.~(\ref{soundspeeds})).

It is now useful to observe that the length of the constriction
introduces a characteristic time scale $t_0 =d/c_2$, the travel time
of an edge wave across the constriction.  This clearly identifies a
short ($t < t_0$) and long ($t>t_0$) time regime.  In these two
regimes, the equations (\ref{sum}) may be approximated by taking the
small and large $d$-limit in the $k$-dependent factor. First we get
for $d\to 0$ the expressions
\begin{equation}
\begin{split}
ReS_-(t,d\to 0)=&e^{-2\theta_{12}}\\ &\times \left[-\frac12
\ln\left(2-2\cos\left(\frac{2\pi c_2t}{L}\right)\right)\right]\\
\simeq & -e^{-2\theta_{12}}\ln(2\pi c_2t/L_2),\\ ImS_-(t,d\to
0)=&e^{-2\theta_{12}}\left(\frac{\pi}2 (1-4 c_2t/L_2)\right)\\ \simeq
&e^{-2\theta_{12}}\frac{\pi}2. \label{sum2}
\end{split}
\end{equation}
In this limit the function $G_-(t)$ will then read \be G_{-,d\to0}(t)=
\left(\delta-\frac{2\pi ic_2t}{L_2}\right)^
{-\frac{2}{\nu}\frac{{e^*}^2}{e^2} e^{-2\theta_1}}.
\label{gdzero}
\ee Remembering that the velocity $c_2$ and the length $L_2$ are
related by Eq.~(\ref{defL2}) we recover exactly the result one has
when the constriction is not present.

In the other limit $d\to \infty$ we have substituted in these two
functions the averaged values, $\langle \cos(k_2d)\rangle=0$, $\langle
\sin^2(k_2d)\rangle=1/2$ obtaining
\begin{equation}
\begin{split}
ReS_-(t,d\to \infty)=&(1+\tanh^2(\theta_{12}))\\ &\times
\left[-\frac12 \ln\left(2-2\cos\left(\frac{2\pi
c_2t}{L_2}\right)\right)\right]\\ \simeq &
-(1+\tanh^2(\theta_{12}))\ln(2\pi c_2t/L_2) ,\\ ImS_-(t,d\to
\infty)=&(1+\tanh^2(\theta_{12}))\left(\frac{\pi}2 (1-2
c_2t/L_2)\right)\\ \simeq &
(1+\tanh^2(\theta_{12}))\frac{\pi}2. \label{sum3}
\end{split}
\end{equation}
Again when we consider the function $G_-(t)$ we get a power law of $t$
\be G_{-,d\to\infty}(t)=\left(\delta-\frac{2\pi
ic_2t}{L_2}\right)^{-\frac{2}{\nu}\frac{{e^*}^2}{e^2} e^{-2\theta_2}
(1+\tanh^2(\theta_{12}))}.
\label{gdinfinity}
\end{equation}
In this case the presence of the constriction affects the exponent of
this correlation function and can change the behavior of the tunneling
amplitude.

The two limiting regimes of short and long times of $S_-(t,d)$ are
clearly visible in the full numerical evaluation as shown in Fig.
\ref{resum.ps}.  In the calculation of the real and
imaginary parts of $S_-(t,d)$, we have fixed a value of $d$ and then
varied the value of $c_2t$. As it is seen in the figures the two
limits we have discussed are reached when $d\ll c_2t$ or $d\gg
c_2t$. In Fig. \ref{resum.ps} we plot the numerical
result for these functions as a function of $c_2t/L$ for some value of
$d$ and fixed $\theta_{12}$. We have restricted the calculation only
to the limit of large system size $ct/L,d/L\ll 1$. The agreement of the
calculated expressions with the approximate results (\ref{sum2}),
(\ref{sum3}) is very good.  Hence from now on we will use the simple
expressions (\ref{sum2}) and (\ref{sum3}) to carry out the calculation
of the tunneling conductance.
\begin{figure}
\begin{center}
\includegraphics[clip,width=8.5cm]{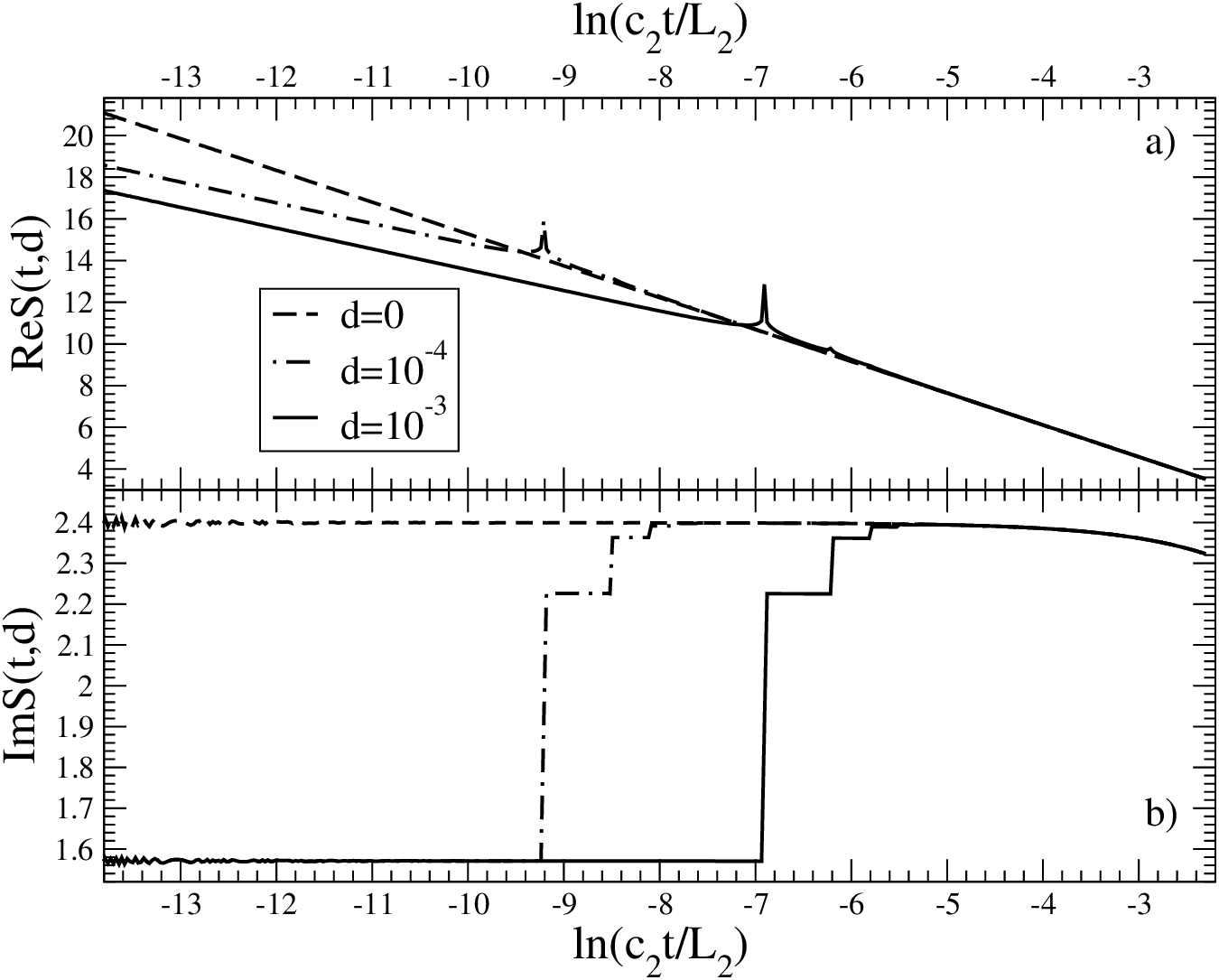}
\caption{{\bf a)} Plot of $ReS_-(t,d)$ vs. $\ln(c_2t/L_2)$ for various values
of $d$. Observe the two different regimes for $c_2t>d$ and $c_2t<d$.
The two slopes agree very well with the approximated result of
Eq.(\ref{Sminus}). We have chosen $\exp(\theta_{12})=1.5275$ in this
calculation.
{\bf b)} Plot of $ImS(t,d)$ vs. $\ln(c_2t/L_2)$ for various values of $d$. We
used the same parameters as a). The values for
small and large $c_2t/L_2$ agree well with the expected results (see
Eqs.(\ref{sum2}) and (\ref{sum3})).  The downward curvature at large
times arises from the finite size of the system used in the numerical
calculation and disappears in the limit of large system size.}
\label{resum.ps}
\end{center}
\end{figure}

%\begin{figure}
%\begin{center}
%\includegraphics[clip,width=7cm]{imsum_new.eps}
%\caption{{\bf Combine this with fig. 4.  Put labels next to curves. }
%Plot of $ImS(t,d)$ vs. $\ln(c_2t/L_2)$ for various values of $d$. We
%used the same parameters as in Fig. \ref{resum.ps}. The values for
%small and large $c_2t/L_2$ agree well with the expected results (see
%Eqs.(\ref{sum2}) and (\ref{sum3})).  The downward curvature at large
%times arises from the finite size of the system used in the numerical
%calculation and disappears in the thermodynamic limit.}
%\label{imsum.ps}
%\end{center}
%\end{figure}
We then approximate the whole sum (\ref{sum}) with a combination of
two functions in the form \be \label{Sminus}
\begin{split}
S_-(t,d)=&\Theta(t-t_0)S_-(t,d\to 0)\\ &+\Theta(t_0-t)(S_-(t,d\to
\infty)-\Delta_-)
\end{split}
\ee where the two functions $S_-(t,d\to 0)$ and $S_-(t,d\to \infty)$
are determined by the corresponding limits for the functions $ReS_-$
and $Im S_- $.  The factor $\Delta_-$ assures that $S_-(t,d)$ is a
continuous function of $t$. With this approximation we have separated
the long time and short time behaviors of the response function. We
then expect that the low energy behavior (which corresponds to the low
bias voltage region) of the conductance will be dominated by the long
time part of $S_-(t,d)$.  Vice-versa, the response to a high bias
voltage will be dominated by the short time behavior of $S_-(t,d)$.
Within this approximation the function $G_-$ reads \be \label{approxG}
\begin{split}
G_-(t) =&\Theta(t-t_0)G_{-,d\to 0}(t)\\ &+\Theta(t_0-t)G_{-,d\to
\infty}(t)\\
&\times\exp\left[-\frac{2{e^*}^2}{e^2\nu}e^{-2\theta_2}\Delta_-\right].
\end{split}
\ee

Having obtained the expression for the function $G_-$ it is now
possible to calculate the response function.  We take into account the finite
potential difference across the Hall bar via the
replacement \be G_-(t)\rightarrow G_-(t)e^{i\frac {e^* V_T
t}{\hbar}} \ee
where 
\be V_T=\frac{V_1+V_2}{2}-\frac{V_3+V_4}{2}= V_H,  
\ee 
is the
potential difference across the quantum point contact, and coincides
with the Hall voltage.
With this transformation, as is shown in the Appendix \ref{mt}, we get
\be
\begin{split}
\label{gtunnelingt0pre}
g_T(\omega_T) =4\frac{|\Gamma|^2
{e^*}^2}{\hbar^3}\frac{d}{d\omega_T}\mathrm{Im} \int_0^\infty
dte^{i\omega_T t}\mathrm{Im}G_-(t).
\end{split}
\ee A lengthy but straightforward calculation gives the expression for
the tunneling conductance at zero temperature (see Appendix
\ref{calculusofg} for details)
\begin{widetext}
\be
\label{gtunnelingt0}
\begin{split}
g_T(\omega_T)=&\left(\frac{4|\Gamma|^2 {e^*}^2 t_0}{\hbar^3}\right)
\left(\frac{a}{2\pi c t_0}\right)^{\alpha}\sin\left(\frac{\pi
\alpha}{2}\right)\\ &\times\frac{d}{d\omega_T}\left[ |\omega_T
t_0|^{\alpha-1}
\left(\cos\left(\frac{\pi\alpha}{2}\right)~\sgn{\omega_T
t_0}\mathrm{Re} \Gamma(1-\alpha,-i \omega_T
t_0)+\sin\left(\frac{\pi\alpha}{2}\right)\mathrm{Im}
\Gamma(1-\alpha,-i\omega_T
t_0)\right)\phantom{\frac{\beta}{2}}\right.\\ &+\left.|\omega_T
t_0|^{\beta-1} \left(\cos\left(\frac{\pi\beta}{2}\right)\sgn{\omega_T
t_0}(\Gamma(1-\beta)-\mathrm{Re}\Gamma(1-\beta,-i\omega_T t_0))
-\sin\left(\frac{\pi\beta}{2}\right)\mathrm{Im}\Gamma(1-\beta,-i\omega_T
t_0)\right)\right]\\
\end{split}
\ee
\end{widetext}
where we have defined \be
\begin{split}
\omega_T=&\frac{e^*}{\hbar}V_T,\\ \alpha=&\frac{2}{\nu}
\frac{{e^*}^2}{e^2}e^{-2\theta_1},\\
\beta=&\frac{2}{\nu}\frac{{e^*}^2}{e^2}
e^{-2\theta_2}(1+\tanh^2(\theta_{12})),
\end{split}\ee
$a$ is a short-distance cut-off,
 and $\Gamma(z_1,z_2)$ is the incomplete $\Gamma$ function
\cite{Abramowitz1964}.

In Fig.~\ref{conductance-t0.eps} we plot $R_{xx}(\omega_T)$ in the
case that the inter-edge interaction is confined to the region of the
constriction (i.e., we set $\theta_1 = \theta_3=0$ and let $\theta_2$
assume several different values).  Experimentally, $\theta_2$ can be
increased by narrowing the constriction by the application of a gate
potential. When $\theta_2 = 0$ there is no interaction and $R_{xx}$
diverges as $V_T^{\alpha-2}$ at low bias.
\begin{figure}[ht]
\begin{center}
\includegraphics[clip,width=6.5cm]{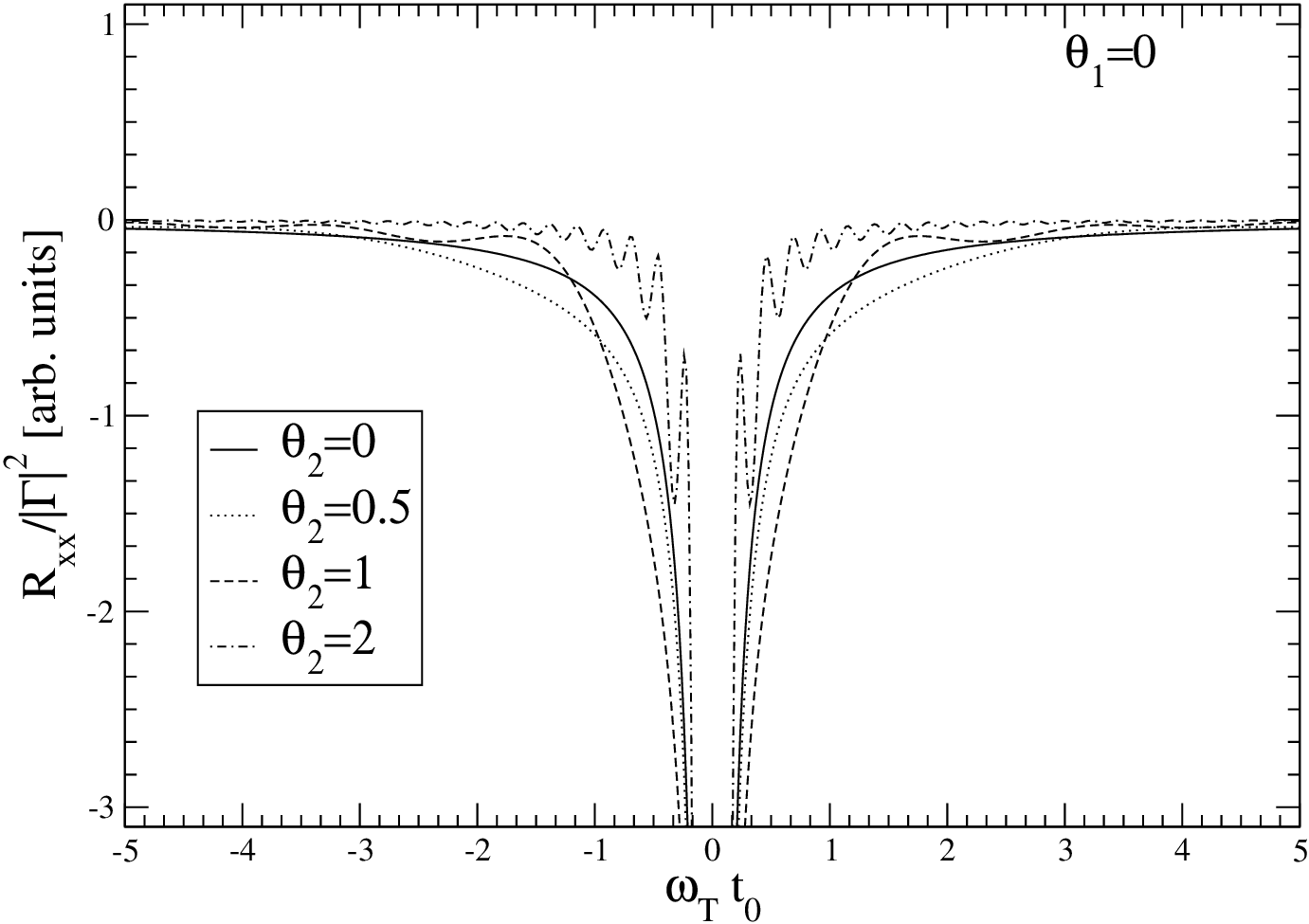}
\caption{Plot of the resistance $R_{xx}/|\Gamma|^2$ given by
Eq. (\ref{tunnel-resistor}) with the $g_T$ calculated in
Eq. (\ref{gtunnelingt0}) for various values of $\theta_2$ at fixed
$\theta_1=0$.  The oscillations at large bias voltage becomes more and
more pronounced with increasing $\theta_2$.  
%{\bf Shall we say that we
%are plotting $R_{xx}/\Gamma^2$?  Notice that perturbation theory
%breaks down when $R_{xx}$ becomes comparable to $\frac{h}{e^2}$-- I
%have mentioned this in theintroduction and at the end of Section VI}
}
\label{conductance-t0.eps}
\end{center}
\end{figure}
This low-bias behavior does not change upon increasing $\theta_2$
because the long time behavior is dominated by the exponent $\alpha$
which does not depend on $\theta_2$.  At larger bias voltage on the other hand, $R_{xx}$ behaves as $V_T^{\beta-2}$. Furthermore,
the plot of $R_{xx}$ shows oscillations, which become more pronounced
with increasing $\theta_2$. We can express the period of these
oscillations in terms of the physical parameters of the theory \be
\Delta V_T=\frac{h}{e^* t_0}=\frac{h c_1}{e^*d}\frac{\cosh
2\theta_1}{\cosh 2\theta_2}~.  \ee The frequency of the oscillations
increases with increasing $\theta_2$ as it is apparent in
Fig. \ref{conductance-t0.eps}.

The finite temperature behavior of the tunneling resistance can be
derived from the zero-temperature behavior of the same quantity by
means of the conformal transformation \cite{Shankar1990} \be
\label{conftransf} \left(\delta\pm i t\right) \to \frac{\sin[\pi
T(\delta\pm it)]}{\pi T}~.  \ee Notice that we are using units in
which $\hbar = k_B=1$, where $k_B$ is the Boltzmann constant.  The
correct physical dimensions are restored via the substitution $T\to
k_B T/\hbar$ and this is understood in the Eqs. (\ref{final-gt}) 
and (\ref{finalgt_wen}) below.  Making the transformation~(\ref{conftransf}) in
Eqs.~(\ref{gdzero}) and ~(\ref{gdinfinity}), and substituting in
Eqs.~(\ref{Sminus}) and (\ref{approxG}) we obtain, after lengthy
calculations (see the Appendix \ref{calculusofg} for more details),
\begin{widetext}
\be
\begin{split}
g_T(\omega_T,T)=&4\frac{{e^*}^2|\Gamma|^2}{\hbar^3}\left(\frac{a}{2\pi
c}\right) \left(\frac{a
T}{2c}\right)^{\alpha-1}\frac{\sin\left(\frac{\pi
\alpha}{2}\right)}{\sinh^\alpha(\pi T t_0)}\frac{\partial}{\partial
\omega_T} \mathrm{Im}\left\{ \frac{e^{i\omega_T
t_0}}{\alpha-i\frac{\omega_T}{\pi T}}
F\left(\alpha,1;1+\frac{\alpha}{2}-i\frac{\omega_T}{2\pi
T};\frac{1}{1-e^{2\pi Tt_0}}\right) \right.\\ &\left.  +2^{\beta-1}
\sinh^\beta(\pi T t_0) B\left(\frac{\beta}{2}-i\frac{\omega_T}{2\pi
T},1-\beta\right) -\frac{e^{i\omega_T t_0}}{\beta-i\frac{\omega_T}{\pi
T}} F\left(\beta,1;1+\frac{\beta}{2}-i\frac{\omega_T}{2\pi
T};\frac{1}{1-e^{2\pi T t_0}}\right) \right\},
\end{split}
\label{final-gt}
\ee
\end{widetext}
where $F$ is the hypergeometric function of four arguments (also
indicated as ${_2}F_1$) and $B$ the Euler beta function
\cite{Abramowitz1964}.  In the case $\theta_1=\theta_2$ we have
$\alpha=\beta$, the first and third term cancel against each other and
we recover Wen's result \be
\begin{split}
g_T(\omega_T)=&4\frac{{e^*}^2|\Gamma|^2}{\hbar^3}\left(\frac{a}{2\pi
c}\right)\left(\frac{a T}{c}
\right)^{\alpha-1}\sin\left(\frac{\alpha\pi}{2}\right) \\
&\times\frac{d}{d\omega_T}\mathrm{Im}
B\left(\frac{\alpha}{2}-i\frac{\omega_T}{2\pi T},1-\alpha\right).
\end{split}
\label{finalgt_wen}
\ee

In Fig. \ref{fig_tnot0.eps}(a-d)
\begin{figure}[ht]
\includegraphics[clip,width=8.5cm]{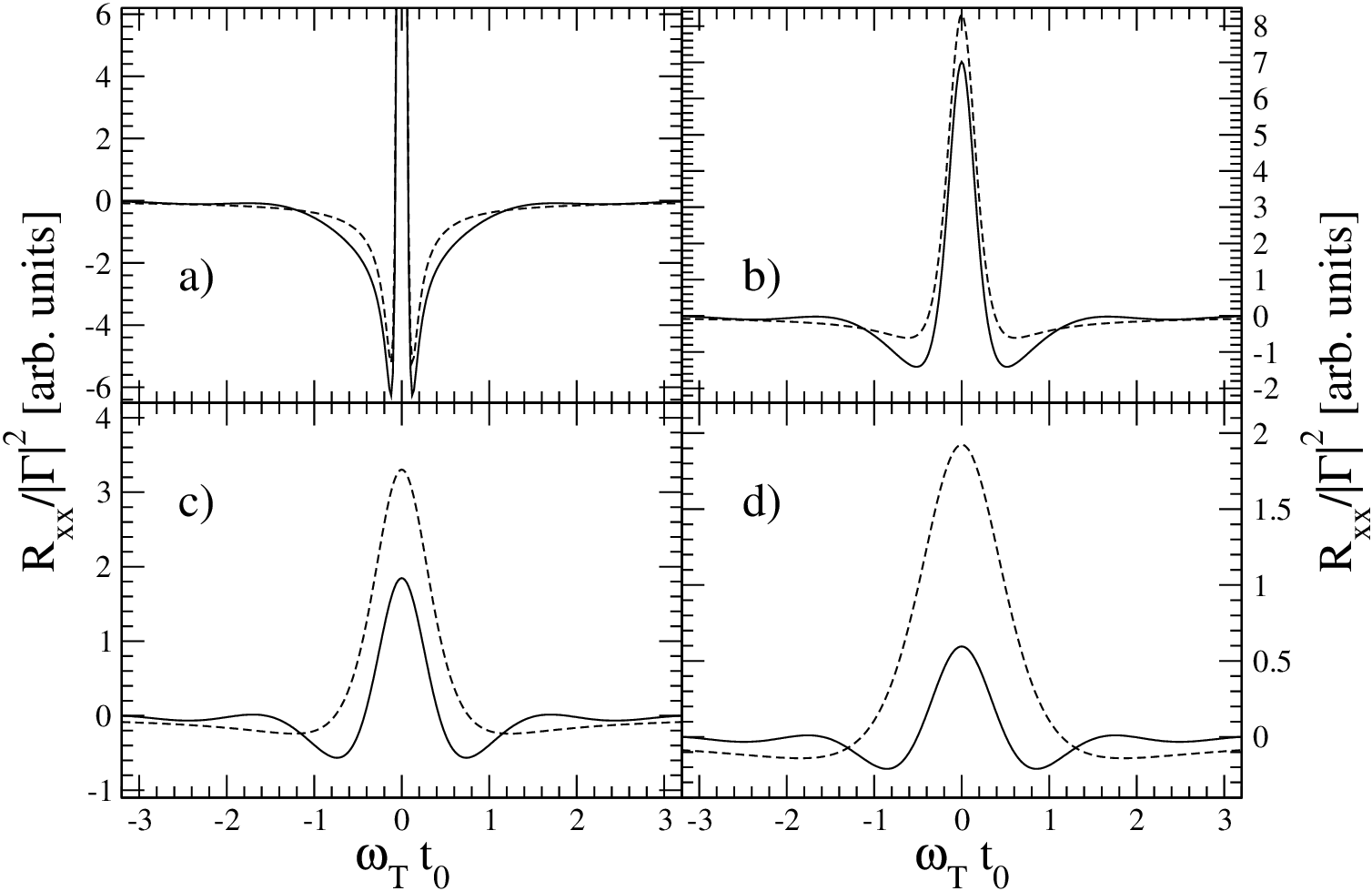}
\caption{Plot of the differential resistance $R_{xx}/|\Gamma|^2$ vs.  $\omega_T$
for a system with inter-edge interaction within the constriction
(continuous line, $\theta_1=0$, $\theta_2=1$) and without inter-edge
interaction (dashed line, $\theta_1=\theta_2=0$).  The four curves
correspond to different temperatures: $\pi T d/c_1=0.1$ (a),
$0.5$ (b), $1$ (c), and $1.5$ (d).}
\label{fig_tnot0.eps}
\end{figure}
we plot the differential resistance $R_{xx}$ vs. bias voltage for a
system {\it without} inter-edge interaction (dashed line --
$\theta_2=\theta_1=0)$ and {\it with} inter-edge interaction (solid
line -- $\theta_1=0$, $\theta_2=1$) for different values of $\pi d
T/c_1 =0.1, 0.5,1$, and $1.5$.  The non vanishing value of $\theta_2$
within the constriction induces oscillations in the $R_{xx}$ vs.
$\omega_T$ relation with the same period as in the zero temperature
case. However, we now have a maximum at zero bias voltage and two
minima at finite bias voltage.  This behavior is due to the fact that
the temperature introduces a new energy scale.  When the $e^*V > k_BT$
we are essentially in the zero temperature case and the resistance
$R_{xx}$ decreases with decreasing bias voltage (see
Fig. \ref{conductance-t0.eps}). But, when the $e^*V<k_BT$ the
resistance turns around and begins to increase, reaching a maximum at
zero bias.  This behavior implies the presence of two minima located
at bias voltages of the order of magnitude of $k_BT/e^*$: these are
clearly seen in Fig. \ref{fig_tnot0.eps}.  The finite value of
$R_{xx}$ at zero bias (independent of $V_T$ to first order) indicates
that the constriction is behaving like an ohmic resistor in this
regime, even though the resistance is strongly temperature-dependent.

The presence of a constriction adds another energy scale in the
problem, associated with the inverse of the characteristic time $t_0$.
For temperatures smaller than $\hbar/t_0$ the low bias behavior is
dominated by the same exponent $\alpha$ (cf. Fig. \ref{fig_tnot0.eps}
(a,b)) irrespective of whether the inter-edge interaction is present
or not.  When the temperature, instead, is greater than $\hbar/t_0$
the exponent $\beta$, which depends on the strength of the interaction
within the constriction, controls the behavior of $R_{xx}$
(cf. Fig. \ref{fig_tnot0.eps}(c,d)). As a consequence the minima at
finite bias are generally deeper and shift to lower voltages.

The effect of the constriction depends quantitatively on both the
inter-edge interaction parameter $\theta_2$ and the temperature.  To
appreciate this we plot in Fig. \ref{fig_tnot02.eps} the differential
resistance $R_{xx}$ for different values of the inter-edge interaction
and the temperature. More specifically we have plotted $R_{xx}$
without interactions ($\theta_2=\theta_1=0)$ and with interactions
within the constriction ($\theta_1=0$, $\theta_2=0.2$) for $\pi d
T/c_1 =0.5, 1, 5$, and $10$.
\begin{figure}[ht]
\includegraphics[clip,width=8.5cm]{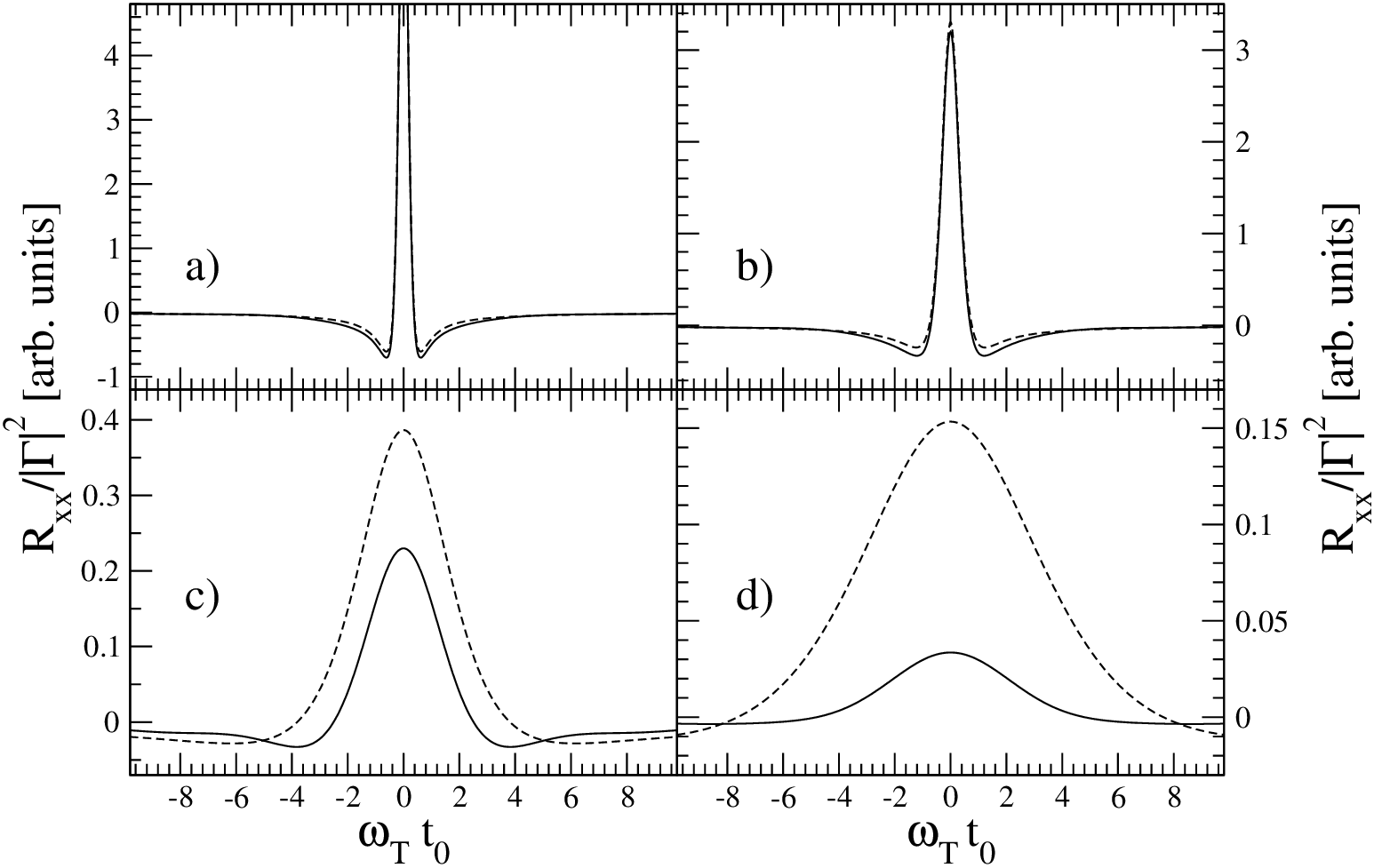}
\caption{Plot of the differential resistance $R_{xx}/|\Gamma|^2$ vs.  frequency
with and without inter-edge interaction within the constriction.
Solid line -- $\theta_1=0$, $\theta_2=0.2$; Dashed line --
$\theta_1=\theta_2=0$.  Temperatures are $\pi d T/c_1 =0.5$ (a),
$1$ (b), $5$ (c), and $10$ (d). }
\label{fig_tnot02.eps}
\end{figure}
We notice that the effect of the inter-edge interaction disappears at
sufficiently low temperature, since it is always the long times
exponent $\alpha$ that matters in that regime.  The effect of the
interaction shows up upon increasing the temperature above the
crossover energy $\hbar/t_0$: the latter decreases with increasing
$\theta_2$.  Such a trend is clearly seen by comparing
Figs. \ref{fig_tnot0.eps} and \ref{fig_tnot02.eps}. We note that 
similar crossover effects in the temperature and voltage behavior 
have been discussed also in the context of transport in quantum
wires \cite{Furusaki1996,Lal2001,Kleinmann2002}.

Finally we would like to comment about recent measurements of
tunneling characteristics through a constriction \cite{Roddaro2002} in
the weak inter-edge tunneling regime at high magnetic field.  At
relatively high temperatures ($T > 400 mK$) the experiment clearly
shows the emergence of a zero bias peak in the differential
longitudinal resistance which is qualitatively consistent with the
results presented above.  The experiment also shows well defined
minima at finite bias voltage, which, according to the previous
discussions may reveal the effect of the constriction. In fact, the
system without inter-edge interactions never shows deep minima in this
temperature range.

At lower temperatures, on the other hand, the experiment shows a
completely different behavior which is not qualitatively consistent
with the present theory irrespective of the presence of inter-edge
interactions. Strong tunneling effects\cite{Fendley1995},
% {\bf cite Fendley and Saleur},
which can be treated by the thermodynamic Bethe Ansatz, are not likely
to explain the unexpected {\it decrease} in $R_{xx}$ that is seen at
these temperatures.  This clearly suggests that a different physical
mechanism comes into play at these temperatures and some additional
physical input is needed. One could, for instance, speculate that,
within the constriction, the hydrodynamic approximation may be too
crude and better treatment of the edge structure may be required.
This is, however, outside the scope of the present work.

%Now we want to discuss the origin of the minima. 
%In that
%theory the presence of the temperature  
%In our model the presence of the zero bias maximum is confirmed. However more 
%evident minima are present. This can be explained by considering that
%in the high bias voltage region the 
%response function has a behavior similar to the zero temperature case but with an exponent $\beta$ which, when $\theta_2\not= 0$, 
%is smaller than $\alpha$ and leading to a stronger negative low-bias divergence. Again when the 
%voltage is below the thermal energy scale the resistance approaches 
%the zero bias positive maximum explaining the presence of the two deeper minima. 
%
%In Fig. \ref{fig_tnot02.eps}
%\begin{figure}[ht]
%\includegraphics[clip,width=7cm]{plot_tnot02.eps}
%\caption{Plot of the differential conductance vs. frequency at fixed mixing
%angle for various temperatures.}
%\label{fig_tnot02.eps}
%\end{figure}
%we plot the differential conductance at $\theta_2 = 1$ and $\theta_1=0$
%for various temperatures. We see from the 
%Fig. \ref{fig_tnot02.eps} that the maximum at $\omega_T=0$ increases when the temperature is 
%lowered. Also the minima strongly depend on the temperature.
%Indeed by lowering the temperature the $R_{xx}$ must approach as limit
% $T\to 0$ the curve reported in Fig. \ref{conductance-t0.eps} hence these
%minima get close and become deeper. 

\section{Conclusion}
In this paper we have extended the derivation of the $\chi$LL model to
arbitrary values of the filling factor $\nu$.  We have developed a
theory to calculate the Landauer-B\"uttiker conductances for various
experimental setups, taking into account both inter-edge tunneling and
inter-edge interactions.  In the absence of tunneling, our model
recovers the usual fractional Hall conductance, even when an
inter-edge interaction is present.  The breaking of translational
invariance, due to the constriction, does not change the low-frequency
behavior of the conductance as long as tunneling can be neglected.

We have then discussed the effect of inter-edge tunneling. Tunneling
destroys the exact quantization of the Hall conductance. We have
calculated the tunneling conductance (related to the resistance of the
constriction) to the second order in the tunneling amplitude. A
problem with the present form of our theory is a fundamental
uncertainty about the value of the effective charge $e^*$ of the
quasiparticles at generic filling factor $\nu$.  This remains a major
open theoretical question.

The presence of the constriction introduces a finite time-scale (the
time it takes an edge wave to travel along the constriction) and gives
rise to different short- and long-time behaviors of the tunneling
propagator.  The long time (low frequency) behavior is dominated by an
exponent that coincides with the one well known in the literature. The
short-time behavior is dominated by a different exponent, smaller than
the long-time exponent.  The interplay between the two exponents
introduces small oscillations in the tunneling conductance which can
possibly be used to measure the amplitude of the inter edge
interaction and the velocity of the modes in the constriction.

\acknowledgments The authors would like to acknowledge discussions
with S. Roddaro and V.~Pellegrini on the subject.  This work was
partially supported by INFM through the PRA 2001-MESODYF project.
G.V. acknowledges partial support from the NSF Grant No. DMR-0074959.

\appendix
\begin{widetext} 
\section{Commutation Relations}\label{app-rel}

To derive the commutation rule of Eq.(\ref{commutation}), one
integrates Eq.(\ref{commutation2}) with respect to $y$ and $y'$
according to the prescriptions given in Eq.(\ref{integrateddensity}).
The delta function makes the commutator non vanishing only for points
belonging to the same edge and the final result will have a factor
$\delta_{\alpha \beta}$.  We get
\begin{eqnarray}
[\delta{\hat \rho}(x_{\alpha}), \delta{\hat \rho}(x_{\alpha}')]&=&
i\ell^2 \int {\rm d}y {\rm d}y' (\partial_{x_\alpha} \rho_0
(x_{\alpha},y))\partial_y
\delta(x_{\alpha}-x_{\alpha}')\delta(y-y')\nonumber\\ &&-i\ell^2 \int
{\rm d}y {\rm d}y' (\partial_y \rho_0
(x_{\alpha},y))\partial_{x_{\alpha}}
\delta(x_{\alpha}-x_{\alpha}')\delta(y-y').
\end{eqnarray}
One observes that the first term on the right-hand side vanishes,
while the second term gives, after making the integration over $y'$
and $y$, for $\alpha =L$
\begin{eqnarray}
[\delta{\hat \rho}(x_{L}), \delta{\hat \rho}(x_{L}')]&=& -i\ell^2 (
\rho_0 (x_{L},d)-\rho_0 (x_{L},0))\partial_{x_{L}}
\delta(x_{L}-x_{L}')\nonumber\\ &=&-i\ell^2 \rho_0(x)\partial_{x_{L}}
\delta(x_{L}-x_{L}')
\end{eqnarray}
and for $\alpha =R$
\begin{eqnarray}
[\delta{\hat \rho}(x_{R}), \delta{\hat \rho}(x_{R}')]&=& -i\ell^2 (
\rho_0 (x_{R},0)-\rho_0 (x_{R},-d))\partial_{x_{R}}
\delta(x_{R}-x_{R}')\nonumber\\ &=& i\ell^2 \rho_0 (x)\partial_{x_{R}}
\delta(x_{R}-x_{R}')
\end{eqnarray}
where $d$ indicates the
distance from the edge at which the density has reached its bulk
value, $\rho_0(x)$.  The different
order of the limits of integration for the two edges gives the
relative minus sign between the edges.
\end{widetext}

\section{Properties of the eigenvalue problem}\label{app-eig}
%{\bf This is fine, it only needs to be checked for accuracy}
In this appendix we want to study some analytical properties of the
equation (\ref{eigenfun}).  First of all let us define the operators
\begin{eqnarray}
M_{\alpha}&=&is_\alpha\partial_{x_\alpha},\\
H_{\alpha,\beta}&=&\frac{\nu}{2\pi}\int_{-\infty}^{\infty}dx_\beta'~
\partial_{x_\alpha}V(x_\alpha,x_\beta')\partial_{x_\beta'}.
\end{eqnarray}
With this definition we rewrite the equation of motion
(\ref{eigenfun}) in the compact form \be \omega M \varphi=H\varphi.
\ee It is easy to see that $H$ and $M$ are hermitian operators and we
request that $H$ is positive definite (this will assure the stability
of the physical system).
%This will assure the stability of the physical system %and we can
%accomplish this by requesting that the diagonal elements of $V$ are
%positive for every value of $x$ and $\det(V)>0$.

Let us define the auxiliary function \be \Psi=H^\frac12\varphi \ee
which is a solution of the equation \be
\frac{1}{\omega}\Psi=\left(H^{-\frac12}MH^{-\frac12}\right)\Psi=\tilde{M}\Psi
\label{new_prob} 
\ee if $\varphi$ is a solution of (\ref{eigenfun}). Because
$\tilde{M}$ is a hermitian operator we have the results:
\begin{enumerate}
\item the set $\{\Psi\}$ of solutions forms a complete base of the
Hilbert space,
\item the orthonormality condition is \be \sum_{\alpha}\int dx_\alpha
\Psi_{n}^*(x_\alpha)\Psi_{m}(x_\alpha)=\delta_{n,m}, \ee
\item the completeness relation \be
\sum_{n}\Psi_{n}(x_\alpha)\Psi_{m}^*(x_\beta')=\delta_{\alpha,\beta}
\delta(x_\alpha-x_\beta').  \ee
\end{enumerate}

Because there is a one-to-one relation between $\varphi$ and $\Psi$ we
have the following properties of the solutions of equation
(\ref{eigenfun}):
\begin{enumerate}
\item the solutions $\varphi$ form a complete base of the Hilbert
space,
\item they are orthogonal with respect to the scalar product \be
(\varphi_n,\varphi_m)=\sum_{\alpha}\int dx_\alpha~ \omega_n
\varphi_{n}^*(x_\alpha)M_{\alpha}\varphi_{m}(x_\alpha), \ee
\item they satisfy the completeness relation \be -i \sum_n \omega_n
\varphi_{n}(x_\alpha)\varphi_{n}^*(x'_\beta)s_\beta\partial_{x_\beta'}=
\delta_{\alpha,\beta}\delta(x_\alpha-x'_\beta).  \ee
\end{enumerate}
We obtain the relations reported in the text if we normalize the
functions $\varphi_n$ as $\varphi_n/\sqrt{|\omega_n|}$.

Now we want discuss the degeneracy of the eigenvalues of the equation
(\ref{eigenfun}):
\begin{itemize}
\item If $\varphi_{m}(x_\alpha)$ is a solution with given eigenvalue
$\omega_m$ then the function $\varphi_{m}^*(x_\alpha)$ is also a
solution with eigenvalue $\omega_{-n}=-\omega_n$.
\item If $\varphi_{m}(x_\alpha)$ is a solution with given eigenvalue
$\omega_m$ then the function $\sigma_{\alpha,\beta}^x
\varphi_{m}(x_\beta)$ is also a solution with eigenvalue $-\omega_m$.
\end{itemize}
Then we have that if $\varphi_{m}(x_\alpha)$ is a solution then
$\sigma_{\alpha,\beta}^x\varphi^*(x_\beta)$ is still a solution with
the same eigenvalue: that is the solutions of problem (\ref{eigenfun})
are doubly degenerate.

%%%%%%%%%%%%%%%%%%%%%%%%%%%%%%%%%%%%%%%%%%%%%%%%%%%%%%%%%
\section{Conservation laws}\label{con_law}
%{\bf  I think this appendix should be reduced to a quick proof of  Eq. (C2).  
%We will then point out in section IV that  this implies $|t|^2+|r|^2 = 1$.  
%There is no need in this paper to go into deeper discussions}

In the case $V_{\alpha,\beta}(x-x')=V_{\alpha,\beta}(x)\delta(x-x')$
the quantity \be
\varphi_\alpha^\dagger(x)\sigma_{\alpha,\beta}^z\varphi_\beta(x)=\varphi^\dagger_L(x)\varphi_L(y)
-\varphi_R^\dagger(x)\varphi_R(x) \label{conserved} \ee is conserved
\be
\partial_x\left(\varphi_\alpha^\dagger(x)\sigma_{\alpha,\beta}^z\varphi_\beta(x)\right)=0.
\label{conserved1} \ee The proof of the existence of this
conserved quantity rests on the basis of the existence of the inverse
of the matrix $V_{\alpha,\beta}(x)$ for every value of $x$.  Consider
the equation of motion and its complex conjugate for the displacement
field wave function $\varphi (x)$
\begin{eqnarray}
i\omega \varphi_{\alpha} (x)& =& \frac{\nu s_{\alpha}}{ 2\pi}
 V_{\alpha\beta}(x)\partial_{x}\varphi_{\beta} (x)\\ -i\omega
 \varphi^{*}_{\alpha} (x)& =& \frac{\nu s_{\alpha}}{ 2\pi}
 \partial_{x}\varphi^{*}_{\beta} (x) V_{\beta\alpha}(x) .
\end{eqnarray} 
The conservation law follows by first taking $V_{\alpha\beta}$ on the
left-hand side of both equations and then multiplying the first
(second) equation on the left (right) by
$\varphi^*_{\beta}\sigma^z_{\beta\alpha}
(\sigma^z_{\alpha\beta}\varphi_{\beta} )$, and finally summing the two
equations.

\section{Calculation of the tunneling propagator}\label{mt}
%{\bf This is fine, but note the proposed change in title.  
%It should also be checked for clarity and accuracy}
In this Appendix we derive Eq.(\ref{gtunneling}). The first task is to
compute $\tilde M_T (t)$. We do it to second order in perturbation
theory in $\Gamma$. The first term in the definition of
$\tilde{M}_T(t)$ %(\ref{mttunneling}) is the tunneling current
propagator. Since it is already second order in $\Gamma$, we only need
to evaluate its average in the unperturbed ground state.  This can be
expressed in terms of the correlation functions $G_-(t';t)$ defined in
(\ref{correlator})
%ausiliary 
%operators $\hat A(x,t)=\Psi^\dagger_L(x,t)\Psi_R(x,t)$ 
obtaining \be
\begin{split}
\left\langle[\hat H_T(t'),\hat H_T(t)]\right\rangle
%=& 2i|\Gamma |^2 {\rm Im} \left\langle[\hat A(x',t'),\hat A^\dagger(x,t)]\right\rangle\\
= 4i |\Gamma|^2 {\rm Im}G_-(t';t)
\end{split}
\ee Notice that in this expression we have dropped the anomalous
averages that appear when one considers the average value of several
field operators $\Psi$.  This is justified by the presence of the
fermion operator $U$ in the definition of the quasi-particle operators
(\ref{psisol}).  Hence the contribution of tunneling current
propagator to $\tilde M_T (t)$ reads \be \left\langle[\hat I_T(t),\hat
I_T(0)]\right\rangle = -4 i\frac{{e^*}^2}{\hbar^2}|\Gamma|^2 {\rm
Im}G_-(t;0).  \ee We now consider the other term in $\tilde M_T
(t)$. This is the average of the tunneling Hamiltonian and is only
first order in $\Gamma$ so that we need to compute the first order
correction to the ground state as well. We get
\begin{eqnarray}
\langle \hat H_T(t)\rangle&=&\frac{i}{\hbar} \int_{-\infty}^t
 dt'~\left\langle[\hat H_T(t'),\hat H_T(t)]\right\rangle\\
 &=&\frac{i}{\hbar} \int_{-\infty}^t dt'~ 4i |\Gamma|^2 {\rm
 Im}G_-(t';t)
\end{eqnarray}
We are now ready to compute the Fourier transform of ${\tilde M}_T
(t)$. We get \be
\begin{split}
\lim_{\omega\to 0} \frac{\tilde M_T(\omega)}{\omega}=&-\frac{4 {e^*}^2
|\Gamma|^2}{\hbar^3}\left[ i\lim_{\omega\to 0}\int_0^\infty dt~
\frac{\sin(\omega t)}{\omega}{\rm Im}G_-(t)\right.\\
&+\left.\lim_{\omega\to 0}\int_0^\infty dt~ \frac{\cos(\omega
t)-1}{\omega}{\rm Im}G_-(t)\right]\\ =&-i\frac{4 {e^*}^2
|\Gamma|^2}{\hbar^3}\int_0^\infty dt~ t{\rm Im}G_-(t).
\end{split}
\ee

The presence of a voltage difference between the edges can be taken in
to account by means of the transformation \be
\begin{split}
G_-(t)\to e^{i\omega_T t}G_-(t),\\ G_+(t)\to e^{i\omega_T t}G_+(t).
\end{split}
\ee from which we get \be
\begin{split}
&\left\langle[\hat H_T(t'),\hat H_T(t)]\right\rangle= 4i
|\Gamma|^2\cos(\omega_Tt) {\rm Im}G_-(t';t),\\ &\left\langle[\hat
I_T(t),\hat I_T(0)]\right\rangle=- 4i\frac{{e^*}^2|\Gamma|^2}{\hbar^2}
\cos(\omega_Tt) {\rm Im}G_-(t)
\end{split}
\ee and we finally arrive to the expression for $ g_T$ \be i
\lim_{\omega\to 0} \frac{\tilde
M_T(\omega)}{\omega}=\frac{4{e^*}^2|\Gamma|^2}{\hbar^3}
\frac{d}{d\omega_T} {\rm Im}\int_0^\infty dt~e^{i\omega_T t}
{\rm Im}G_-(t).  \ee

\section{Evaluation of integrals}\label{calculusofg}
%{\bf This is fine, but note the proposed change in title. 
% It should also be checked for clarity and accuracy}
In this appendix we provide a few details concerning the evaluation of
the integral occurring in the calculation of the Fourier transform of
the response function $G_-(t)$.  In the zero temperature case, we must
evaluate an integral of the form 
\be
\label{integral1}
\int_{t_0}^\infty dt (\delta\pm it)^{-\alpha} e^{i\omega t} 
\ee
where $\alpha$ is a positive real number. To do this we go in the
complex plane of the variable $t$ and consider, for positive frequency
$\omega$, an integration path as the one shown in
Fig. \ref{integrationpath.eps}.  A specular path in the lower half
plane must be used for negative frequency.  We observe that the
integrand function has no poles in the complex half-plane of $t$ with
a positive real part. Hence the integral on the whole path is
zero. The integral on the arc vanishes by letting the radius go to
infinity.
\begin{figure}
\includegraphics[clip,width=7cm]{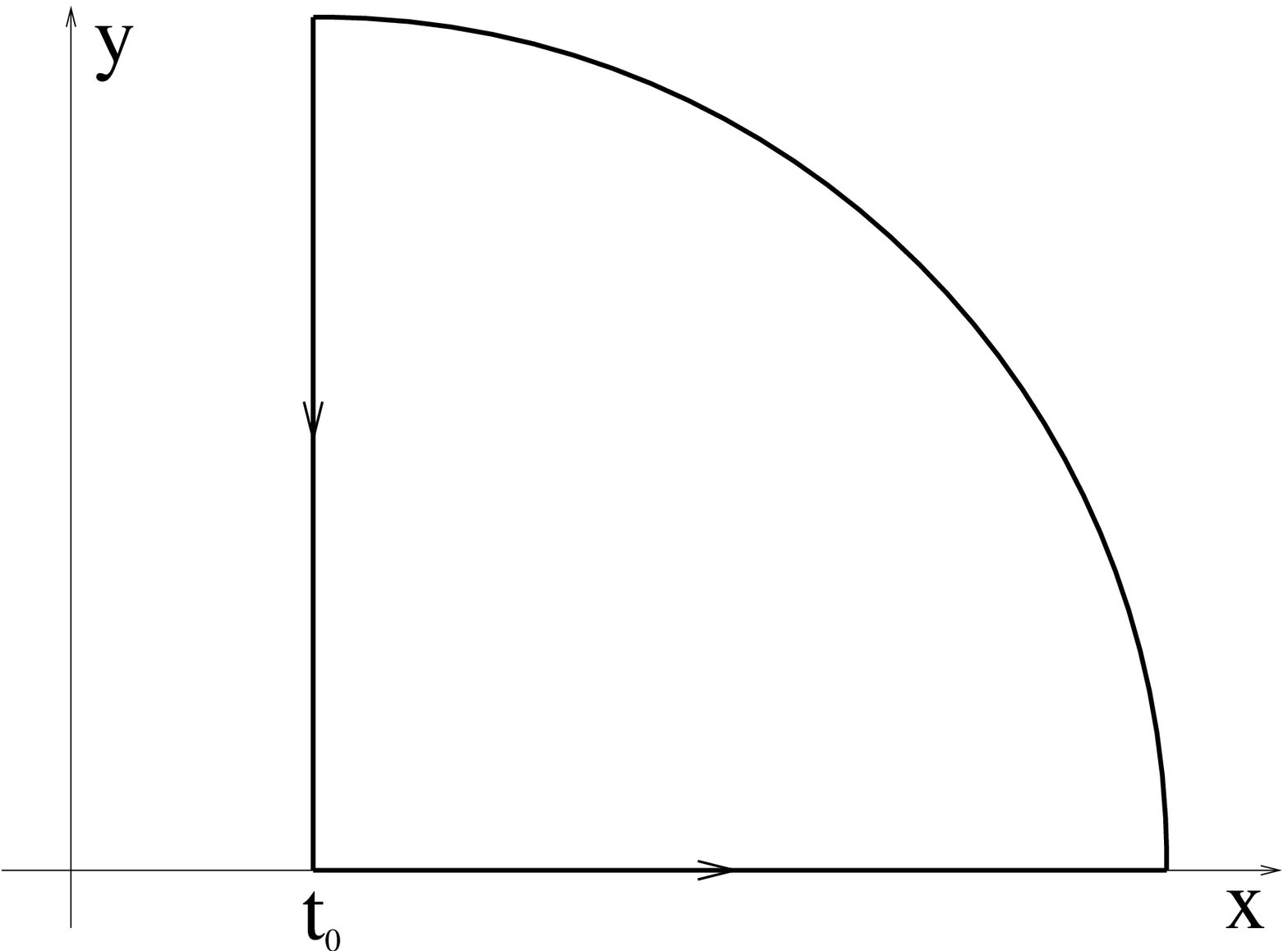}
\caption{The integration path used to evaluate the integral
(\ref{integral1}) when the frequency is positive. A similar path,
closed in the lower plane, is used in the case $\omega<0$.}
\label{integrationpath.eps}
\end{figure}

As a result we get 
\be \int_{t_0}^\infty dt (\delta\pm
it)^{-\alpha} e^{i\omega t}\stackrel{\delta \to 0}{=} i (\mp
1)^{-\alpha} \omega^{\alpha-1} \Gamma(1-\alpha,-i\omega t_0).  
\ee 
The
case $t_0=0$ can be carried out by calculating the convolution product
between the Fourier transform of the $\Theta(t)$ function and the
integral \be
\begin{split}
\int_{-\infty}^\infty dt (\delta\pm it)^{-\alpha} e^{i\omega
t}=&2 \sin(\pi\alpha) e^{i\frac{\pi}{2}\alpha}|\omega|^{-1-\alpha}\\
&\times \Gamma(1+\alpha)(\mp 1)^{-1-\alpha},
\end{split}
\ee obtained by cutting the complex plane along the imaginary axis,
starting from $t=\pm i\delta$.

In the finite temperature case, we need to calculate the integral 
\be
\int_{t_0}^\infty dt~e^{i\omega t}\frac{{(\pi
T)}^\alpha}{\sinh(\pi T t)^\alpha}.  
\ee One can easily obtain the
result reported in the text by means of the substitution $s=e^{-2\pi
Tt}$ which reduces the above integral to the definition of the
hypergeometric $F$ function of four
arguments\cite{Abramowitz1964,Gradshteyn1965}. The integral with $t_0=
0$ can be easily obtained in the limit $t_0\to 0$ by using the
corresponding limiting expression of the hypergeometric function $F$
in terms of the Euler beta function.

%%%%%%%%%%%%%%%%%%%%%%%%%%%%%%Here starts the bibliography%%%%%%%%%%%%%%%%%%%%%%%%%%%%%%%%%%%
\bibliography{qhe-biblio}
\end{document}